\documentclass{IEEEtaes}

\usepackage{color,array,amsthm}
\usepackage{graphicx}

\usepackage{amssymb}
\usepackage{amsmath}
\usepackage{flushend}

\usepackage{cleveref}

\usepackage{mathtools}
\usepackage{etoolbox}

\makeatletter

\def\ps@headings{%
  \def\@oddhead{}%
  \def\@evenhead{}%
  \def\@oddfoot{\hfil\normalfont\footnotesize\thepage\hfil}%
  \let\@evenfoot\@oddfoot
}

\def\ps@plain{%
  \def\@oddhead{}%
  \def\@evenhead{}%
  \def\@oddfoot{\hfil\normalfont\footnotesize\thepage\hfil}%
  \let\@evenfoot\@oddfoot
}

\makeatother

\DeclareMathOperator*{\argmin}{arg\,min}

\newcommand{\norm}[1]{\left\lVert#1\right\rVert}
\newtheorem{problem}{Problem Definition}

\begin{document}

\title{HYDRA: Quantifying Botnet Resource Thresholds for Efficient Link-Flooding Attacks on LEO Satellite Networks}

\author{Roee Idan}
\author{Rami Puzis}
\author{Asaf Shabtai}
\author{Yuval Elovici}

\affil{Stein Faculty of Computer and Information Science, Ben-Gurion University of the Negev, Beer-Sheva, Israel}

\corresp{{\itshape (Corresponding author: R. Idan, roeeidan@post.bgu.ac.il).}}

\authoraddress{Roee Idan, Rami Puzis, Asaf Shabtai, and Yuval Elovici are with the Stein Faculty of Computer and Information Science, Ben-Gurion University of the Negev, Beer-Sheva, Israel (e-mail: roeeidan@post.bgu.ac.il; puzis@bgu.ac.il; shabtaia@bgu.ac.il; elovici@bgu.ac.il).}

\markboth{IDAN ET AL.}{HYDRA}

\maketitle

\begin{abstract}

Low Earth orbit (LEO) satellite constellations, such as Starlink and Kuiper, are rapidly emerging as a critical backbone for low-latency global connectivity.
As these systems expand, they become more attractive attack targets, necessitating increased resilience and security.
Threat actors seek to exploit constellation-specific properties such as predictable motion, time-varying topologies, and reliance on inter-satellite and ground-satellite links.
Recent work has shown that link-flooding attacks (LFAs) can exploit these properties to congest strategic network bottlenecks.
Yet, prior work does not quantify the resilience of LEO networks to targeted disruption.
We present HYDRA, a modeling and optimization framework that formulates LFA variants as botnet minimization problems.
HYDRA quantifies network resilience to LFAs by measuring the smallest active subset of bots and the corresponding traffic allocation required to disrupt communication between targeted geographic areas.
Under matched stealth constraints, HYDRA achieves the same targeted disruption as ICARUS while using 34\% fewer bots and 23\% less aggregate attack traffic.
HYDRA achieves over 97\% success in sustaining continuous attacks as the network topology evolves.
Finally, HYDRA evaluates five mitigation strategies, showing how routing diversification, ingress policing, distance-based traffic constraints, source throttling, and botnet attrition reduce attack success and improve network resilience to targeted disruption.

\end{abstract}


\section{Introduction} \label{sec:intro}

Low Earth orbit (LEO) satellite networks, such as SpaceX's \textit{Starlink}~\cite{fcc_starlink1} and Amazon's \textit{Project Kuiper}~\cite{fcc_kuiper}, are transforming global communications.
These constellations consist of thousands of satellites operating at altitudes of up to 2,000 km, forming a new connectivity backbone in space.
They offer low-latency, high-speed Internet access worldwide, extend connectivity to underserved regions, enable direct satellite-to-cell connectivity~\cite{qu2017leo,vatalaro1995analysis,starlinkdirecttocell}, and support applications such as broadband services, financial data transfer, and emergency communications~\cite{bhattacherjee2019network,handley2018delay}.

Modern LEO satellite networks are connected by a mesh of inter-satellite links (ISLs) and ground-satellite links (GSLs)~\cite{giuliari2021icarus}.
While many constellations still route traffic through ground stations (GSs) and terrestrial networks~\cite{deng2025time}, there is growing interest in using ISLs for direct in-network routing.
ISL based routing can create a more self contained space network, reducing dependence on ground infrastructure and enabling more direct end-to-end paths.
Prior studies show that such architectures can halve end-to-end latency compared to terrestrial routing due to faster signal propagation, shorter routes, and fewer hops~\cite{bhattacherjee2019network}.
As a result, end-to-end communication that relies on ISLs for in-network routing is expected to become more common, increasing the performance and flexibility of LEO connectivity~\cite{tesmanian_starlink_afghanistan,hauri2020internet}.

Despite their large scale, LEO constellations are topologically sparse and capacity constrained.
Each satellite typically maintains only a small number of active ISLs and GSLs, and these links have finite capacity~\cite{giuliari2021icarus}.
LEO satellite motion is also highly predictable; publicly available two-line element (TLE) data enables forecasting satellite positions with deviations of only a few kilometers~\cite{celestrak_tle,kelso2007validation}.
As a result, the network topology evolves in a structured and largely foreseeable manner over time.
These characteristics distinguish LEO constellations from terrestrial networks and shape both how traffic is routed and how performance bottlenecks emerge across geographic regions.

The rapid growth of LEO satellite networks has drawn the attention of threat actors, as the number of satellites and the range of offered services increase the exposed surface for cyberattacks.
Researchers in academia and industry have also begun systematically exploring broader cybersecurity risks in LEO constellations.
Prior studies document threats ranging from physical-layer attacks, such as jamming, spoofing, and energy draining~\cite{space_threats,zhang2023energy}, to terminal-level exploits, such as bypassing secure boot protections to access and inspect internals of the Starlink user terminal~\cite{starlink_dish_hack}.
These threats are no longer just theoretical.
In May 2022, the pro-Russian group Killnet claimed responsibility for DDoS attacks on Starlink~\cite{killnet_attack}.
These events, along with others discussed in Section~\ref{sec:background}, highlight adversaries' growing interest in targeting satellite Internet services.
As satellite Internet services expand, these threats motivate a need to better understand the feasibility and resource requirements of disruptive network-layer attacks.

Among the cyber threats LEO constellations are facing, distributed denial-of-service (DDoS) attacks, and in particular link-flooding attacks (LFAs), are especially well suited to exploiting the unique characteristics of these networks.
DDoS attacks use a distributed set of compromised hosts, called a botnet, that are remotely controlled by an adversary to inject traffic and overwhelm limited network resources, while LFAs steer this traffic to congest a small set of strategically chosen links~\cite{studer2009coremelt}.
In LEO networks, an adversary can exploit the global reach of satellite access together with publicly available orbital information to plan targeted congestion at specific times and locations.
This can increase the impact of each active bot by focusing traffic on time-varying bottlenecks, lowering the botnet resources required for targeted disruption.
In this work, we focus on zone attacks, where the adversary uses multiple LFAs to disrupt connectivity between two targeted geographic areas rather than aiming for a network-wide outage or a single endpoint.

Prior work has shown that LFAs against LEO constellations are possible, including snapshot-based demonstrations, attacks that remain effective without assuming exact routing knowledge, attacks that account for short-term topology changes, and attack structures that aim for broader disruption beyond a single precisely targeted path~\cite{giuliari2021icarus,lu2024dosat,wang2024starmaze}.
Collectively, these studies show the potential of targeted congestion attacks in LEO networks.

However, prior work does not address how feasible such attacks are under a constrained population of compromised satellite users.
This question is especially important in LEO networks, where the pool of plausibly usable bots is far smaller than in terrestrial settings: IoT deployments include tens of billions of connected devices~\cite{fagan2023trusted}, whereas the largest LEO constellation, Starlink, currently serves more than 10 million active users~\cite{starlink_10m_2026}.
Since only a subset of satellite users could plausibly be compromised, and each compromised terminal can contribute only uplink capacity on the order of tens of Mbps~\cite{starlinkuploadlimit}, the required botnet size becomes a key measure of attack practicality.
Furthermore, existing work does not quantify attacker efficiency for zone attacks, including how to minimize the number of active bots needed to disrupt connectivity between selected areas, how this requirement changes as the topology evolves over time, and how it scales when the attacker pursues broader or multiple disruption objectives.
It also remains unclear how versatile a single botnet configuration can be across different disruption goals.
Aggregate attack flow alone does not characterize LFA feasibility, because generating and steering a given traffic volume through the targeted bottlenecks depends on the number, uplink capacity, and geographic location of bots.
The minimum botnet required to disrupt connectivity between targeted zones captures the resources an adversary must actually obtain and coordinate, and therefore provides a direct measure of the network's resilience to LFAs: a network that requires a larger botnet to achieve disruption is more resilient.
This leaves a broader security question open: how resilient are LEO networks to LFAs?

To address these questions, we introduce HYDRA, a modeling and optimization framework for LFAs in dynamic LEO networks.
Rather than focusing only on attack execution in a single snapshot, HYDRA models adversarial planning, including bot selection, traffic allocation, and persistence under time-varying topologies.
Given a pool of bots mapped to geographic locations, HYDRA computes the smallest active bot subset and corresponding traffic plan needed to induce targeted congestion under constellation capacity constraints.
Under idealized, fully optimized conditions, HYDRA estimates the minimum botnet resources required to achieve targeted disruption, establishing a baseline for assessing network resilience to LFAs.
HYDRA evaluates this threshold across multiple LFA objectives, from single-snapshot attacks to persistent attacks over time, and examines whether a single botnet can support flexible targeting and simultaneous disruption across multiple targets.
Finally, HYDRA evaluates five mitigation strategies, showing how they reduce attack success and increase the minimum botnet resources required for targeted disruption.

Using HYDRA, we find that the resilience of LEO networks to LFAs is substantially lower than prior attack models imply.
In single-snapshot zone-attack instances, HYDRA reduces the active botnet size by 34\% relative to an ICARUS stealth-distribution baseline under matched detectability constraints, while also reducing aggregate attack flow by 23\%.
HYDRA maintains over 97\% success in continuous attacks under the modeled routing and traffic assumptions, and the required resources scale efficiently when extending disruption to multiple targets.
We further use HYDRA to evaluate five mitigation strategies, showing how changes to routing, ingress capacity, source rate limits, traffic constraints, and bot availability reduce global attack success and increase the minimum botnet resources required for targeted disruption.

The main contributions of this paper are:
\begin{itemize}
\item We model the dynamic topology of LEO satellite networks and define disruption objectives for LFA-based zone attacks.
\item We develop a discrete optimization framework that quantifies resilience to LFAs through the minimum botnet resources and corresponding traffic allocation required to induce targeted congestion under constellation capacity constraints.
\item We evaluate persistent disruption, showing that a single optimized botnet can sustain attacks across time-varying topologies.
\item We study flexible botnet utilization, showing how a single botnet can be reused to target different disruption objectives without reconfiguration.
\item We quantify simultaneous multi-target disruption, showing that a coordinated botnet can attack multiple zone pairs at once more efficiently than planning each target independently.
\item We use HYDRA to evaluate five mitigation strategies, quantifying how changes to routing, ingress capacity, source rate limits, traffic constraints, and bot availability affect global attack success and the minimum botnet resources required for targeted disruption.
\end{itemize}

\section{Background} \label{sec:background}

\subsection{LEO Satellite Network Architecture}

LEO satellite constellations consist of thousands of satellites orbiting Earth at altitudes of up to 2,000 km~\cite{qu2017leo,fcc_starlink1,fcc_starlink2}, enabling low latency and fast data transmission.
However, due to this closeness each satellite has a limited coverage area.
Each LEO satellite completes an orbit in approximately 90 minutes~\cite{fischer2008topology,bhattacherjee2019network}, leading to constant motion and frequent changes in position relative to the Earth's surface.

Due to this rapid movement and narrow coverage area, consistent regional coverage requires many satellites, with handovers every few minutes~\cite{deng2021ultra,bhattacherjee2019network}.
Precise orbital configurations and sufficient satellite density are necessary to ensure uninterrupted global coverage~\cite{vatalaro1995analysis}.
LEO constellations are often arranged in structured patterns, such as the Walker-Delta configuration~\cite{fischer2008topology,bhattacherjee2019network}, across multiple orbital planes to optimize coverage and connectivity.
With sufficient satellite density, this design supports continuous coverage by keeping at least one satellite within communication range of any point on Earth, enabling reliable global connectivity~\cite{bhattacherjee2019network}.

Each satellite is equipped with GSLs that connect to user terminals (UTs) and ground stations (GSes), which in turn connect to the terrestrial internet, and typically supports four ISLs; together, the GSLs and ISLs form a dynamic mesh~\cite{mynaric_space,zhang2022enabling}.
These links transmit data using either radio frequency or optical lasers, with optical ISLs offering higher bandwidth and lower latency~\cite{chaudhry2022crossover}.
However, their capacity remains limited: ISLs typically support tens of Gbps, while GSLs operate at just a few Gbps~\cite{mynaric_space}.

As satellites orbit and the Earth rotates, the constellation topology continually changes~\cite{fischer2008topology}.
ISL connections are reconfigured dynamically, and GSL connections shift with satellite movement to serve different users.
These changes introduce routing variability, resulting in a highly dynamic and complex network structure~\cite{zhang2022enabling}.
Routing in such networks typically relies on topology aware algorithms that adapt to the dynamic connectivity to select low-latency paths~\cite{handley2018delay}.

There is extensive public data on satellite constellations, including orbital elements, launch information, constellation configurations, and community tracking data published by governments, researchers, and hobbyists.
TLE sets, which are freely available and updated every few days~\cite{celestrak_tle}, provide orbital parameters used with models like SGP4~\cite{kelso2007validation} to predict satellite positions with deviations of just 1–2~km.
While useful for legitimate applications, TLEs can be exploited by adversaries to predict satellite locations and reconstruct network topologies~\cite{giuliari2021icarus}, facilitating cyberattack planning.

\subsection{Distributed Denial-of-Service and Link-Flooding Attacks in Terrestrial Networks}

DDoS attacks are a major threat in computer networking, aiming to make resources unavailable to legitimate users~\cite{douligeris2004ddos}.
Attackers often rely on botnets to flood specific servers or infrastructure with large volumes of traffic~\cite{antonakakis2017understanding}.
A notable example is the Mirai botnet~\cite{antonakakis2017understanding}, which infected hundreds of thousands of IoT devices and used them to launch massive DDoS attacks.

In addition to traditional high-volume flooding attacks, more sophisticated strategies use low-rate flows to exploit network topology and congest targeted links.
In the \textit{Coremelt} attack~\cite{studer2009coremelt}, bots exchange traffic in a way that causes it to converge on core ISP links without directly targeting end systems.
The \textit{Crossfire} attack~\cite{kang2013crossfire} refines this approach by sending traffic to public decoy servers, carefully selected so that the traffic paths traverse critical links. 
While these flows appear benign, when coordinated within a botnet they can cause sustained congestion.  
Both attacks demonstrate how knowledge of routing and topology can enable stealthy, persistent disruption without relying on high traffic volumes.

\subsection{Cybersecurity Challenges in Space and Satellite Networks}
The growing reliance on space-based infrastructure introduces unique cybersecurity challenges.
Space systems operate in harsh environments and face limitations such as constrained computational resources, long development cycles, and difficult to deploy software updates~\cite{willbold2023space,yue2023low}.

Attacks targeting space systems include jamming, spoofing, malware injection, data interception, and DoS attacks~\cite{yue2023low,willbold2023space}.
Among these threats, Willbold et al.~\cite{willbold2023space} performed an experimental security analysis of real satellite firmware and discovered multiple critical vulnerabilities, showing that software and firmware weaknesses can be exploited to compromise satellite systems and achieve persistent control.

In February 2022, a cyberattack on Viasat's KA-SAT network disrupted broadband across Ukraine and Europe, affecting thousands of users and critical infrastructure~\cite{viasat_case}.
Viasat traced the attack to a misconfigured VPN appliance, which illustrates how ground-side vulnerabilities can compromise satellite operations~\cite{viasat2022update}.

In May 2022, the pro-Russian group \textit{Killnet} claimed responsibility for DDoS attacks targeting SpaceX's Starlink satellite internet services~\cite{killnet_attack}.
This incident highlights the increased targeting of satellite networks, particularly in conflict zones.

Beyond disruption, satellite links can also be abused as part of an adversary's operational infrastructure.
The Turla APT group used satellite connectivity to support cyber-espionage activity against government and military organizations~\cite{kaspersky_turla,symantec_turla}.
By routing traffic through satellite connections and hijacking IP address space, Turla increased its anonymity and made detection more difficult~\cite{kaspersky_turla,symantec_turla}.

Together, these incidents reflect adversaries' growing interest in the space sector and highlight that satellite connectivity can be disrupted or abused through both ground-side compromise and network-layer attacks.

\section{Related Work: LFAs on LEO Satellite Networks} \label{sec:related_work}

While the principles of LFAs were developed for terrestrial networks, recent research has begun to adapt them for the unique, dynamic environment of LEO constellations.

ICARUS~\cite{giuliari2021icarus} was the pioneering attack.
It provided the first dedicated framework for simulating LFAs against LEO networks and successfully demonstrated that an adversary could congest specific ISLs and GSLs by exploiting the network's predictable topology and routing.
This work was foundational in establishing the general feasibility of the threat.

Subsequent research built upon this foundation and explored more dynamic attack strategies.
The DoSat study~\cite{lu2024dosat} showed that attacks could be timed to coincide with vulnerable link handover periods, a moment of weakness in the network, to maximize their disruptive impact.
The StarMaze study~\cite{wang2024starmaze} introduced a novel attack variant that creates persistent traffic loops within the constellation's regular, ring-like topologies, focusing on network-wide disruption by targeting entire satellite rings and causing more widespread damage.

While the state-of-the-art works demonstrate the feasibility of LFAs and their importance and impact on LEO networks, the studies share common limitations that motivate our research.
As can be seen in Table~\ref{tab:compare}, which summarizes prior work, researchers have primarily focused on execution of the attack, assuming the existence of the necessary attack infrastructure while overlooking the following aspects:
\begin{itemize}
    \item \textbf{Botnet optimization:} Prior work assumes access to a large, well-distributed botnet and does not address how to minimize the number of bots required for an attack under per-bot upload constraints.
    This overlooks a key limitation in LEO networks: the number of user terminals that could serve as bots is limited, and each bot can inject only a limited amount of attack traffic.
    
    \item \textbf{Long-term persistence and adaptation strategy:} While some studies examined short-term dynamics, most prior work did not formalize or optimize the problem of sustaining a targeted attack against an inter-zone connection over extended periods across a dynamic satellite topology.
    
    \item \textbf{Multi-target efficiency:} Prior work focused on disrupting single links or single zones.
    The studies did not explore whether an adversary could efficiently use a single botnet to simultaneously disrupt communication between multiple, distinct pairs of geographic zones.
    
    \item \textbf{Mitigation impact on attack thresholds:} Existing studies offer limited evaluation of how mitigations affect the botnet resources required for disruption in dynamic LEO topologies.
\end{itemize}

HYDRA addresses these underexplored aspects by explicitly modeling the strategic planning phase of an attack.
It quantifies resilience to LFAs through the minimum botnet resources required for targeted disruption, evaluates this threshold across persistent, flexible, and simultaneous attack objectives, and assesses how mitigations affect both attack feasibility and the threshold. 

\begin{table*}[hbtp]
\centering
\caption{Comparison of HYDRA to state-of-the-art LFA approaches for LEO networks.}\label{tab:compare}
\begin{tabular}{|l|c|c|c|c|}
\hline
\textbf{Aspects Considered} & \textbf{HYDRA} & \textbf{ICARUS} \cite{giuliari2021icarus} & \textbf{DoSat} \cite{lu2024dosat} & \textbf{StarMaze} \cite{wang2024starmaze} \\ \hline
\textbf{Targeting ISLs} & \checkmark & \checkmark & \checkmark & \checkmark \\ \hline
\textbf{Snapshot-Based Network Modeling} & \checkmark & \checkmark & \checkmark & \checkmark \\ \hline
\textbf{Adaptation to Dynamic Network Topologies} & \checkmark & & \checkmark & \checkmark \\ \hline
\textbf{Continuous Attacks} & \checkmark & & \checkmark & \checkmark \\ \hline
\textbf{Long-Lasting Attack Strategy} & \checkmark & & & \checkmark \\ \hline
\textbf{Active Botnet Minimization} & \checkmark & & & \\ \hline
\textbf{Flexible Botnet Utilization} & \checkmark & & & \\ \hline
\textbf{Simultaneous Attacks on Multiple Zone Pairs} & \checkmark & & & \\ \hline
\textbf{Defense / Mitigation Evaluation} & \checkmark & \checkmark & & \\ \hline
\end{tabular}
\end{table*}
\section{The HYDRA Attack} \label{sec:sysmodel}
In this section, we define the HYDRA attack model and its optimization-based mathematical formulation for zone attacks in LEO satellite networks.
We present the threat model, formal attack definitions, optimization-based planning methods, and execution phase constraints.  

\subsection{High-Level Attack Overview}

The HYDRA attack unfolds in two phases:

\begin{itemize}
    \item \textbf{Weaponization Phase:} The attacker determines the optimal layout and composition of the botnet. The goal is to minimize the number of bots while ensuring that sufficient traffic can be generated to disrupt communication between the targeted areas.
    \item \textbf{Execution Phase:} Once the botnet is assembled, the attacker executes the attack by directing carefully constructed traffic flows to congest critical links without violating background capacity constraints.
\end{itemize}

\begin{figure}[hbt]
\centering
\includegraphics[width=\columnwidth]{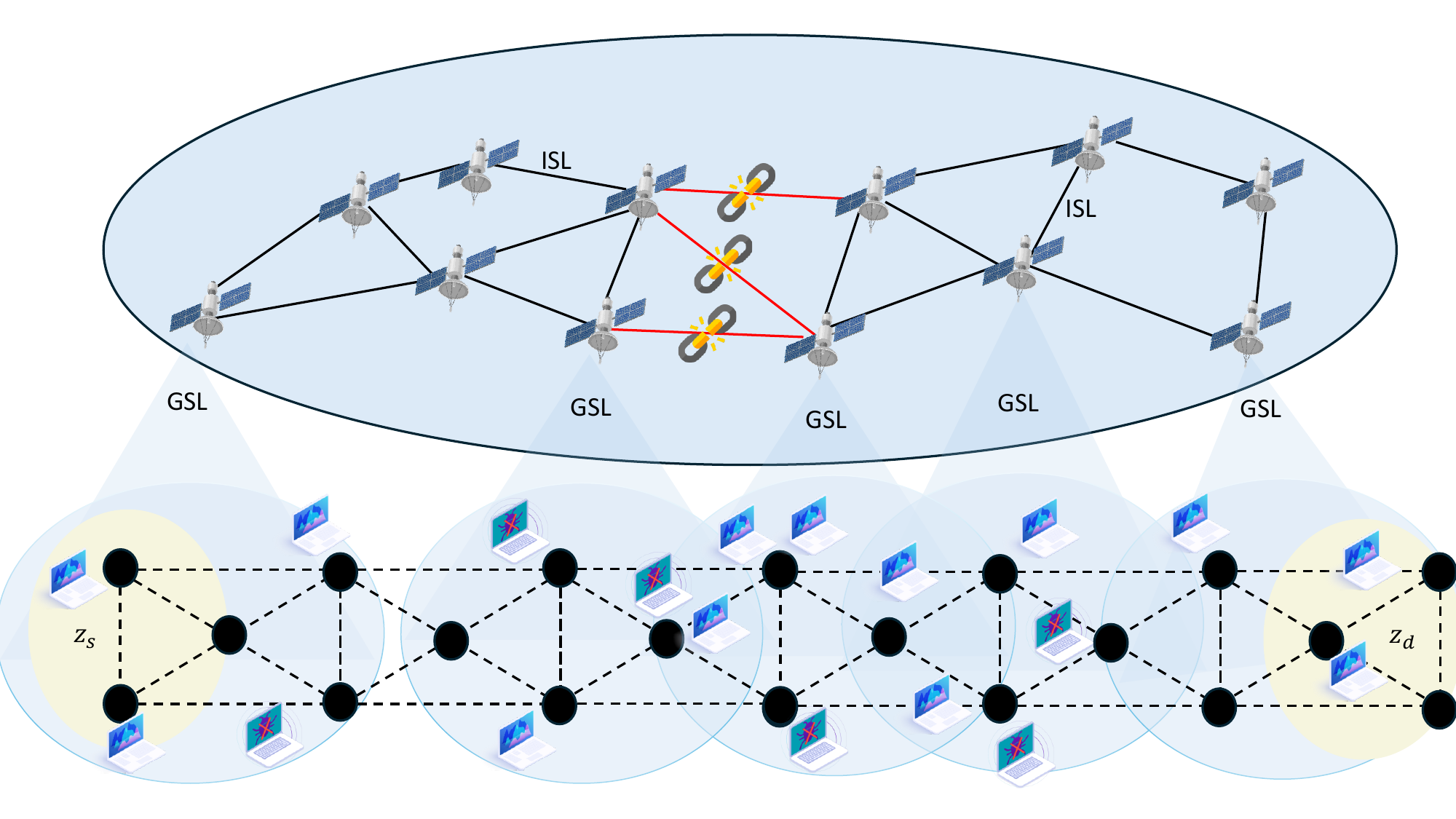}
\caption{Illustration of the HYDRA attack during a specific snapshot. Attack traffic is injected into the LEO satellite constellation via ground points of origin (GPOs) and flows through GSLs and ISLs toward its destination. The goal is to congest selected links (highlighted in red), disrupting communication between a source zone ($z_s$) and a destination zone ($z_d$) based on the topology at that snapshot.}
\label{fig:forpaper}
\end{figure}

This targeted congestion effectively disrupts the communication paths between the designated areas, as illustrated in Figure~\ref{fig:forpaper}.

\subsection{Threat Model}

\subsubsection{Adversary’s Capabilities}  
We assume an adversary with detailed knowledge of the constellation topology and satellite orbits, based on publicly available sources such as NORAD~\cite{celestrak_tle,kelso2007validation}.  
This enables accurate prediction of satellite positions over time, which is valuable for attack planning.
The adversary can analyze observable traffic from compromised endpoints to infer the routing policy and obtain coarse estimates of link capacities and background loads.
They also control a globally distributed botnet and can synchronize its activity to converge traffic on targeted links simultaneously~\cite{antonakakis2017understanding,rasti2015temporal}.

\subsubsection{Attacker's Objective}
The adversary's objective is to identify the smallest subset of the botnet and corresponding data flows needed to disrupt communication between targeted geographic zones, while maintaining a persistent attack over time.

\subsection{Preliminaries: The Satellite Network}
\subsubsection{Dynamic Network Topology} \label{sec:network overview}
We model the LEO network as a time-varying graph ($G_t$) captured at periodic time snapshots $t \in T$.
The network comprises thousands of satellites ($N^{S}$) and ground point origins (GPOs), denoted as $N^{GPO}$. 
GPOs represent discrete locations on Earth based on a geodesic grid~\cite{sahr2003geodesic}. 
We assume that $N^{S}$ and $N^{GPO}$ remain fixed over the time scales we study, and thus we do not model changes in them.
However, satellite positions change continuously, and link connectivity changes with them.
Satellites communicate with each other via ISLs ($L^{ISL}_t\subseteq N^{S}\times N^{S} $) and with the ground via GSLs ($L^{GSL}_t\subseteq N^{S}\times N^{GPO}$). 
The LEO network at time $t$ is represented as $G_t=(N^{S}\cup N^{GPO}, L^{GSL}_t \cup L^{ISL}_t)$. 
In the remainder of the paper, we may omit the time index $t$ when the context makes it clear, to simplify notation.

\subsubsection{\texorpdfstring{Traffic Matrix ($\mathbf{TM}$)}{Traffic Matrix (TM)}}\label{sec:def:tm}
In our simulation traffic flows between pairs of GPOs: a source $g_s \in N^{GPO}$ and a destination $g_d \in N^{GPO}$.
The amount of traffic sent by $g_s$ to $g_d$ at every time point $t$ is captured by the traffic matrix $\mathbf{TM}_{t,g_{s},g_{d}}$.

\subsubsection{\texorpdfstring{Routing Function ($\mathcal{F}$)}{Routing Function (F)}}
We define a routing function $\mathcal{F}_t(g_s,g_d)$ which maps 
a source-destination pair ($g_s,g_d$)
to a sequence of links (a path) in  $G_t$.
We assume that shortest-path routing is used to maintain reliable communication~\cite{handley2018delay}.
For simplicity, and to keep the formulation focused on botnet minimization rather than routing dynamics, we abstract routing as a single route per $(g_s,g_d)$ pair at each snapshot (i.e., no path dispersion).
This separation keeps HYDRA focused on quantifying attack resource requirements for a given routing policy, while alternative routing policies are evaluated separately as mitigation mechanisms in \Cref{sec:mitigations}.

\subsubsection{Residual Capacity}\label{sec:def:capacity}
Each link $l \in L^{ISL}\cup L^{GSL}$ has a maximum capacity $c_{l_\text{max}}$.
Link usage is the total traffic exchanged between GPOs through the link.
The residual capacity is the remaining bandwidth on link $l$ at time $t$ after accounting for all traffic traversing the link.
We consider a link congested if its residual capacity drops below a fraction $(1-\alpha)$ of its maximum capacity, i.e.,
$(1-\alpha)\cdot c_{l_\text{max}}$ .
The parameter $\alpha \in (0,1)$ sets the residual-capacity margin used to classify links as congested, providing a common congestion criterion across snapshots, zone pairs, and attack configurations.
When a link is congested, it may result in increased latency, packet loss, or dropped connections between communicating nodes~\cite{studer2009coremelt}.

\subsection{The HYDRA Attack Model}\label{sec:MathRepr}
In this section, we formulate the constraints and functions that define the attack and describe how the network and attacker’s actions are modeled, how the injected traffic is represented, and how we ensure the network capacity constraints.

\subsubsection{\texorpdfstring{Attack Traffic Matrix ($\mathbf{ATM}$)}{Attack Traffic Matrix (ATM)}}
The attack traffic matrix $\mathbf{ATM}$ captures the additional traffic sent by the botnet as part of the attack.
Each entry $\mathbf{ATM}_{t,g_s,g_d}$ denotes the amount of attack traffic sent from source GPO $g_s$ to destination GPO $g_d$ at snapshot $t$.

\subsubsection{Attack Target} 
Consider a geographic zone such as a country, state, or region. 
We define a zone as a subset of GPOs $z\subseteq N^{GPO}$.
The adversary's goal is to disrupt communication between one or more pairs of zones, $A = \{(z_1, z_2), \dots\ : z_i\subseteq N^{GPO}\}$. 
We define a function $\mathcal{P}(t, zp)$ that returns the set of all paths between a pair of zones $zp = (z_1, z_2) \in A$ at time $t$ as
\begin{equation}
\mathcal{P}(t, zp) = \left\{ \mathcal{F}_t(g_s, g_d) : g_s \in z_1, g_d \in z_2 \right\}.
\end{equation} 
Since $\mathcal{F}_t(g_s,g_d)$ returns a single route for each GPO pair $(g_s,g_d)$ at snapshot $t$, $\mathcal{P}(t,zp)$ contains one path per pair and thus scales as $|\mathcal{P}(t,zp)| = |z_1||z_2|$, rather than with the number of arbitrary simple paths in the topology.
The attacker's objective is to congest at least one link in every route in $\mathcal{P}(t, zp)$. 
For a specific zone pair $zp = (z_1, z_2) \in A$, we denote $\mathbf{ATM}^{zp}$ as the attack traffic matrix targeting that pair.  
More generally, $\mathbf{ATM}^{A}$ is the set of all such matrices for zone pairs in $A$.

\subsubsection{\texorpdfstring{Post-Attack Available Capacity ($\mathcal{E}$)}{Post-Attack Available Capacity (E)}}

This function represents the remaining available capacity of a link $l$ at time $t$ after accounting for both nominal and attacker-injected traffic.
Formally, it is defined as
\begin{equation}
\mathcal{E}_t(l,\mathbf{ATM}) =  c_{l_\text{max}}
-
\sum_{\substack{g_s,g_d: \\ l \in \mathcal{F}_t(g_s,g_d)}}
(\mathbf{TM}+\mathbf{ATM})_{t,g_s,g_d}.
\end{equation}
We define the residual-capacity congestion threshold of link $l$ as
\begin{equation}
\tau_l = (1-\alpha)c_{l_\text{max}}.
\end{equation}
A link is considered congested when its post-attack residual capacity is at most $\tau_l$ while remaining physically feasible, i.e.,
\begin{equation}
0 \le \mathcal{E}_t(l,\mathbf{ATM}) \le \tau_l .
\end{equation}

\subsubsection{\texorpdfstring{Congestion Condition ($\mathbb{C}$)}{Congestion Condition (C)}}
To disrupt communication, the adversary must ensure that each path $p \in \mathcal{P}(t, zp)$ is congested.
A path is considered congested when it contains at least one link whose aggregate traffic reaches or surpasses the congestion threshold.
This condition evaluates whether all paths between a zone pair $zp$ are congested at time $t$ and can be written as
\begin{equation}
\begin{aligned}
    \mathbb{C}(t, zp, \mathbf{ATM}_t) =
    &\ \forall\, p \in \mathcal{P}(t, zp),\; \exists\, l \in p \text{ s.t.} \\
    &\ 0 \le \mathcal{E}_t(l, \mathbf{ATM}^{zp}_t) \le \tau_l .
\end{aligned}
\end{equation}
An attack is considered successful if $\mathbb{C}(t, zp, \mathbf{ATM}_t)$ is true for all $zp \in A$ and all $t \in T$.
To encode $\mathbb{C}(t,zp,\mathbf{ATM}_t)$ in the optimization problems, we use binary variables indicating whether a link is selected as congested at a snapshot, and enforce that for every path the sum of these binaries over its links is at least one.
Let $L^{\mathrm{sel}}_t(zp,\mathbf{ATM}^{zp}_t) \subseteq L_t$ denote the set of links selected as congested for zone pair $zp$ at time $t$.

\subsubsection{\texorpdfstring{Non-Target Congestion Constraint ($\mathbb{U}$)}{Non-Target Congestion Constraint (U)}}\label{sec:def:U}

To limit congestion to the links selected for the attack, all non-selected links need to remain uncongested.
This condition evaluates whether all non-selected links remain uncongested at time $t$ and can be written as
\begin{equation}
\begin{aligned}
    \mathbb{U}(t, zp, \mathbf{ATM}^{zp}_t) =
    &\ \forall\, l \in L_t \setminus L^{\mathrm{sel}}_t(zp,\mathbf{ATM}^{zp}_t), \\
    &\ \mathcal{E}_t(l,\mathbf{ATM}^{zp}_t) > \tau_l .
\end{aligned}
\end{equation}
An attack satisfies the non-target congestion constraint if $\mathbb{U}(t,zp,\mathbf{ATM}^{zp}_t)$ is true for all $zp \in A$ and all $t \in T$.
To encode $\mathbb{U}(t,zp,\mathbf{ATM}^{zp}_t)$ in the optimization problems, we use the same binary variables that indicate whether a link is selected as congested, and enforce that every non-selected link remains above the congestion threshold.
The goal of this constraint is to help the attacker avoid congestion on links other than the selected target links under the modeled traffic conditions, so that attack traffic can reach the intended congestion points without disrupting itself earlier in the path.

\subsubsection{Total Outgoing Attack Traffic}
Let $\beta$ be the maximum upload capacity of a single host.
We use a uniform per-bot rate limit $\beta$ to keep the formulation focused on botnet minimization; this can be interpreted as a conservative bound (e.g., a lower-percentile upload rate) across the bot population.
We define $\mathcal{S}(g_s, t, \mathbf{ATM})$ as the minimal number of bots the attacker must operate at $g_s$ at time $t$ as
\begin{equation}
\mathcal{S}(g_s, t, \mathbf{ATM}) = \left\lceil \frac{\sum_{g_d} \mathbf{ATM}_{t,g_s,g_d}}{\beta} \right\rceil.
\end{equation}

Let $\mathcal{I}(\mathbf{ATM}^{A}_T)$ denote the set of all source GPOs $g_s$ that appear as a row index in any attack traffic matrix $\mathbf{ATM}^{zp}_t$ for some $zp \in A$ and $t \in T$.  
We define $\mathcal{S}_M(T, A, \mathbf{ATM})$ as a vector indexed by $g_s \in \mathcal{I}(\mathbf{ATM}^{A}_T)$, where each entry represents the maximum number of bots the attacker must operate at $g_s$ across all relevant snapshots and zone pairs
\begin{multline}
\mathcal{S}_M(T, A, \mathbf{ATM}) \\ =
\left[
\max_{\substack{t \in T \\ zp \in A}}
\mathcal{S}(g_s, t, \mathbf{ATM}^{zp}_t)
\right]_{g_s \in \mathcal{I}(\mathbf{ATM}^{A}_T)}.
\end{multline}

\subsection{Weaponization Phase Optimization}\label{sec:attack opt}
During the weaponization phase, the attacker seeks to minimize the number of active bots required to disrupt communication between the targeted zones.
Botnet size is expressed as the $L_1$ norm of
\begin{equation}\label{eq:A star}
    \mathbf{ATM}^* = \argmin_{\mathbf{ATM}_{T}^A} \norm{\mathcal{S}_M(T, A, \mathbf{ATM})}_1.
\end{equation}

For a given disruption objective and set of modeled constraints, the optimum of Eq.~\eqref{eq:A star} identifies the minimum active botnet required to induce targeted congestion and its corresponding traffic allocation.
We use this minimum botnet and corresponding traffic allocation as an operational measure of resilience to LFAs: the more bots a network requires an adversary to deploy to achieve disruption, the greater its resilience.
We aim to fulfill this objective in several attack scenarios (described below), which we formulate as botnet minimization problems (BMPs).
All BMP instances are formulated as mixed-integer linear programs (MILPs) and solved using the Gurobi Optimizer \cite{gurobi}.

\subsubsection{Snapshot Attack Method}\label{sec:SAM}
The snapshot attack method aims to disrupt communication between a single targeted zone pair $zp = (z_1, z_2)$ at a specific snapshot time $t$, while minimizing the botnet size.
The associated single snapshot botnet minimization problem (SS-BMP) is defined as follows

\begin{problem}[SS-BMP]
Given: 
\begin{itemize}
    \setlength\itemsep{0em}
    \item[] $\mathcal{P}(t, zp)$: Set of paths connecting zones $z_1$ and $z_2$ at time $t$.
    \item[] $L_t$: Set of ISLs and GSLs in the network $G_t$ at time $t$.
    \item[] $\mathcal{S}_M(t, zp, \mathbf{ATM})$: Vector of required bots per GPO for the snapshot $(t, zp)$.
    \item[] $\mathbb{C}(t, zp, \mathbf{ATM}^{zp}_t)$: Congestion indicator for all paths in $\mathcal{P}(t, zp)$.
    \item[] $L^{\mathrm{sel}}_t(zp,\mathbf{ATM}^{zp}_t)$: Set of links selected as congested for zone pair $zp$ at time$t$.
    \item[] $\mathbb{U}(t, zp, \mathbf{ATM}^{zp}_t)$: Non-selected link congestion constraint.
\end{itemize}

\noindent Find $\mathbf{ATM}^{zp}_t$: Attack traffic for zone pair $zp$ at time $t$.
\begin{subequations} \label{eq:SAM opt problem}
\begin{align}
  \mathbf{ATM}^{*} = &\argmin_{\mathbf{ATM}^{zp}_t} \norm{\mathcal{S}_M(t, zp, \mathbf{ATM})}_1 \\[1ex]
  \text{s.t.} \hspace*{2em} 
  & \mathbb{C}(t, zp, \mathbf{ATM}^{zp}_t), \label{eq:SAM cong} \\
  & \mathbb{U}(t, zp, \mathbf{ATM}^{zp}_t). \label{eq:SAM non_target_cong}
\end{align}
\end{subequations}
\end{problem}

Constraint \cref{eq:SAM cong} guarantees that each path between the targeted zones is disrupted via at least one congested link.  
Constraint \cref{eq:SAM non_target_cong} ensures that links not selected for congestion remain uncongested.  
Together, these conditions ensure that the attack achieves its goal without obstructing its own execution.

The SS-BMP serves as a baseline for more advanced attacks, laying the groundwork for strategies like sustained attacks over time or simultaneous multi-zone disruption.

\subsubsection{Continuous Attack}\label{sec:cont attack}

We extend the attack over time by optimizing the botnet strategy across multiple snapshots.
Each snapshot captures the network state at a specific time, with the attack aiming to maintain congestion across relevant paths throughout all snapshots in $T$.

For each snapshot $t \in T$, the parameters $\mathcal{P}(t,\ldots)$, $L_t$, $\mathcal{E}_t^{\mathbf{TM}}$, and $\mathbb{C}(t,\ldots)$ vary based on satellite positions, connectivity, and nominal data transfers through the network.  
The overall objective is to minimize the size of the botnet needed to sustain the attack across all snapshots.  
This is done by determining, for each GPO, the maximum number of bots that need to be sent across all $t \in T$, summing these peak values across all GPOs, and minimizing the total number of bots required to maintain the attack throughout the time window.

The associated continuous botnet minimization problem (C-BMP)  is defined as follows
\begin{problem}[C-BMP]
Given:
\begin{itemize}
    \setlength\itemsep{0em}
    \item[] All inputs of SS-BMP.
    \item[] $T$: Set of snapshot times capturing the dynamic topology of the network.
    \item[] $\mathcal{S}_M(T, zp, \mathbf{ATM})$: Vector representing the maximum number of bots needed per GPO across all $t \in T$.
\end{itemize}

\noindent Find $\{\mathbf{ATM}^{zp}_t\}_{t \in T}$: A set of attack traffic matrices for zone pair $zp$ across all snapshots.

\begin{subequations} \label{eq:Continuous opt problem}
\begin{align}
   \mathbf{ATM}^* = &\argmin_{\{\mathbf{ATM}^{zp}_t\}_{t \in T}}
   \norm{\mathcal{S}_M(T, zp, \mathbf{ATM})}_1 \\[1ex]
   \text{s.t.} \hspace*{2em}
   &\forall t \in T,\;\; \mathbb{C}(t, zp, \mathbf{ATM}^{zp}_t), \label{eq:path cong Continuous} \\
    & \forall t \in T,\;\; \mathbb{U}(t, zp, \mathbf{ATM}^{zp}_t). \label{eq:non target cong Continuous}
\end{align}
\end{subequations}
\end{problem}

Solving the optimization problem across all $t \in T$ minimizes the total size of the botnet while maintaining congestion throughout the entire attack window.
This approach adapts to changes in satellite positions, connectivity, and nominal data transfers, ensuring sustained disruption with minimal resources.

\subsubsection{Flexible Botnet Configuration for Targeted Attacks}\label{sec:felxible}

This method aims to identify the smallest botnet that can individually attack any pair of zones within a defined set.
This approach ensures that any zone pair in the set can be targeted without adjusting the botnet.
The optimization here focuses on solving the problem globally to find a minimal set that can be reused for attacking each zone pair individually, for each pair $zp \in A$, where $A$ is the set of all attacked zone pairs.

The associated flexible botnet minimization problem (F-BMP)  is defined as follows
\begin{problem}[F-BMP]
Given:
\begin{itemize}
    \setlength\itemsep{0em}
    \item[] All inputs of C-BMP.
    \item[] $A$: Set of zone pairs that the botnet needs to be able to attack individually.
    \item[] $\mathcal{S}_M(T, A, \mathbf{ATM})$: Vector representing the number of bots required per GPO for each zone pair in $A$ across snapshots.
\end{itemize}

\noindent Find $\{\mathbf{ATM}^{zp}_t\}_{zp \in A,\; t \in T}$: A set of attack traffic matrices for each zone pair $zp$ across all snapshots.

\begin{subequations} \label{eq:Flexible opt problem}
\begin{align}
   \mathbf{ATM}^* = &\argmin_{\{\mathbf{ATM}^{zp}_t\}_{zp \in A,\; t \in T}}
   \norm{\mathcal{S}_M(T, A, \mathbf{ATM})}_1 \\[1ex]
   \text{s.t.} \hspace*{2em}
   &\forall zp \in A,\;\; \forall t \in T,\;\; \mathbb{C}(t, zp, \mathbf{ATM}^{zp}_t), 
   \label{eq:path cong Flexible} \\
   &\forall zp \in A,\;\; \forall t \in T,\;\; \mathbb{U}(t, zp, \mathbf{ATM}^{zp}_t).
\end{align}
\end{subequations}
\end{problem}

Minimizing $\norm{\mathcal{S}_{M}(T, A, \mathbf{ATM})}_1$ yields the smallest botnet capable of executing each attack independently.
Constraint~\eqref{eq:path cong Flexible} ensures that every zone pair $zp \in A$ can be successfully targeted.
Solving the optimization problem across all zone pairs enables the creation of a single botnet configuration that supports all attacks in $A$, improving resource efficiency and allowing flexible targeting without reconfiguration.

\subsubsection{Maximized Simultaneous Attack Method}\label{sec:Simul attack}
This method targets multiple zone pairs simultaneously using a unified set of constraints that span all relevant paths.  
As previously done, the set of targeted zone pairs is denoted as $A$.  
For each snapshot $t \in T$, we define the unified path set $\mathcal{P}_A(t)$ as the union of all communication paths between zone pairs in $A$ during that snapshot  $\mathcal{P}_A(t) = \bigcup_{zp \in A} \mathcal{P}(t, zp)$.
This unified set enables the construction of a single attack traffic matrix $\mathbf{ATM}_t$ for each snapshot, which must simultaneously congest all relevant paths across all zone pairs in $A$.

The associated simultaneous botnet minimization problem (SIMU-BMP)  is defined as follows:
\begin{problem}[SIMU-BMP]
Given:
\begin{itemize}
    \setlength\itemsep{0em}
    \item[] All inputs of SS-BMP.
    \item[] $\mathcal{P}_A(t) = \bigcup_{zp \in A} \mathcal{P}(t, zp)$: Set of all communication paths across the zone pairs in $A$ at snapshot $t$.
\end{itemize}

\noindent Find $\{\mathbf{ATM}_t\}_{t \in T}$: A set of attacker-injected traffic matrices, one per snapshot.

\begin{subequations} \label{eq:SimultOpt}
\begin{align}
   \mathbf{ATM}^* = &\argmin_{\{\mathbf{ATM}_t\}_{t \in T}}
   \norm{\mathcal{S}_M(T, A, \mathbf{ATM})}_1 \\[1ex]
   \text{s.t.} \hspace*{2em}
   &\forall t \in T,\;\; \mathbb{C}(t, A, \mathbf{ATM}_t), \label{eq:path cong Optimal} \\
   &\forall t \in T,\;\; \mathbb{U}(t, A, \mathbf{ATM}_t).
\end{align}
\end{subequations}
\end{problem}

This optimization approach minimizes the botnet size while maximizing the attack effectiveness by identifying shared ISLs and data flows that can congest multiple paths concurrently.

\subsection{Execution Phase and Stealth Constraints}
In the \textit{execution phase}, the optimization is constrained by a pre-existing botnet $B$, ensuring that the total attack traffic sent from any GPO does not exceed the aggregate bandwidth of the bots available at that location. This constraint is enforced via
\begin{equation}
\forall t \in T,\;\; \forall g \in N^{GPO},\;\; \beta \cdot B_g \geq \sum_{g_d} \mathbf{ATM}_{t, g, g_d},
\end{equation}
where $\beta$ is the maximum sending capacity of a single bot, and $B_g$ denotes the number of bots assigned to GPO $g$.
The summation on the right-hand side represents the total attack traffic sent from $g$ at time~$t$.

To further reduce detectability, the attacker can impose additional execution constraints inspired by~\cite{giuliari2021icarus}.
These constraints can restrict the selected congestion targets to ISLs, or limit the amount of attack traffic that traverses any GSL.

For a link $l$, let
\begin{equation}
\mathcal{D}_t(l) =
\left\{(g_s,g_d) : l \in \mathcal{F}_t(g_s,g_d)\right\}
\end{equation}
denote the set of source-destination GPO pairs whose route at time $t$ traverses $l$.
The total attack traffic traversing each GSL can then be bounded by an absolute threshold $u(l)$,
\begin{equation}
    \forall t \in T,\;\; \forall l \in L^{GSL}_t,\;\;
    \sum_{(g_s,g_d)\in \mathcal{D}_t(l)}
    \mathbf{ATM}_{t,g_s,g_d}
    \leq u(l).
\end{equation}
Alternatively, the bound can be defined relative to the nominal residual capacity on the GSL.
We define the pre-attack residual capacity of link $l$ at time $t$ as
\begin{equation}
\phi_t(l) =
c_{l_\text{max}}
-
\sum_{(g_s,g_d)\in \mathcal{D}_t(l)}
\mathbf{TM}_{t,g_s,g_d}.
\end{equation}

The total attack traffic traversing each GSL can then be limited to a fraction $\gamma(l)$ of this residual capacity,
\begin{equation}
    \forall t \in T, \forall l \in L^{GSL}_t,\hspace*{-1.5em}
    \sum_{(g_s,g_d)\in \mathcal{D}_t(l)}
    \hspace*{-1em}\mathbf{ATM}_{t,g_s,g_d}
    \leq \gamma(l)\phi_t(l).
\end{equation}
Additionally, a limit $m(g)$ can be placed on the number of bots used per GPO
\begin{equation}
    \forall g \in N^{GPO},\;\; B_g \leq m(g).
\end{equation}
The term \(B_g\) captures the required bot availability at each GPO, while \(m(g)\) can be used to impose deployment-specific limits, such as population-based or operator-specific caps on the number of bots available at that location.

\section{Evaluation} \label{sec:evaluation}
To evaluate these attack strategies, we simulated 10 constellation formations, including Starlink~\cite{fcc_starlink1, fcc_starlink2}, OneWeb~\cite{onewebsatellite}, and Project Kuiper~\cite{fcc_kuiper}.  
Each setup was based on key parameters: the number of orbital planes, number of satellites per plane, altitude, inclination, and link capacity.  
We evaluated the first Starlink shell (72 planes × 22 satellites, 550~km altitude, 53$^\circ$ inclination).  
Each simulation modeled a 90-hour period with routing updates every 30 seconds, capturing a dynamic topology.
We used an extended ICARUS simulator~\cite{giuliari2021icarus} on a system with 256 AMD EPYC 7763 CPU cores and 1TB RAM.  
The targeted pairs were selected via GDP-weighted sampling of global city pairs to reflect economically important regions.  
The GDP data came from public sources~\cite{spatialecon_gecon,global_cities_gdp}.  
Ten thousand random pairs were examined to assess robustness and scalability.

Link capacities are fixed per link type for ISLs and GSLs.
At each snapshot $t$, the background traffic is generated by sampling source--destination GPO pairs using GDP-weighted GPO weights and assigning each sampled pair to a shortest-path route in $G_t$.
In our evaluation, we use 250,000 source--destination pairs per snapshot, randomly sampled according to the GDP-weighted GPO distribution.
In our experiments, $\alpha=0.9$, corresponding to 90\% utilization and 10\% headroom.
Table~\ref{tab:eval_params} summarizes the concrete values used in our experiments.

The GPOs were represented as discussed in \Cref{sec:network overview}, using a triangulated grid that created 1,800 points.
The per-host upload limit was assumed to be 25~Mbps~\cite{starlinkuploadlimit}.

\begin{table}[t]
\centering
\small
\setlength{\tabcolsep}{5pt}
\caption{Key parameters used in the evaluation.}
\label{tab:eval_params}
\begin{tabular}{|p{0.44\columnwidth}|p{0.48\columnwidth}|}
\hline
\textbf{Parameter} & \textbf{Value} \\
\hline
Snapshot interval & 30 seconds \\
Evaluation window & 90 hours \\
Number of GPOs & 1{,}800 \\
\hline
Background pair sampling & GDP-weighted \\
Sampled background demands per snapshot & $250{,}000$ sampled $(g_s,g_d)$ pairs \\
Routing model & Shortest-path routing \\
\hline
ISL capacity $c_{l_{\max}}$ & 20{,}000~Mbps \\
GSL capacity $c_{l_{\max}}$ & 4{,}000~Mbps \\
\hline
Congestion threshold & $\alpha=0.9$ \\
Per-bot upload cap & $\beta=25$~Mbps \\
\hline
\end{tabular}
\end{table}

\subsection{Botnet Resource-Threshold Comparison}

To compare the botnet resource thresholds estimated by HYDRA and ICARUS, we evaluated single-snapshot attack instances under matched experimental conditions, including the same network snapshot topology, background traffic state, routing model, link capacities, congestion threshold $\alpha$, per-bot upload cap $\beta$, and GPO grid.
DoSat and StarMaze study complementary LEO LFA mechanisms, but we compare them qualitatively in Table~\ref{tab:compare} because we did not identify public implementations that could be faithfully integrated into our evaluation setting.
Because ICARUS emphasizes stealth by distributing attack traffic across uplinks, quantified by the maxUp metric~\cite{giuliari2021icarus}, we constrain HYDRA using the same detectability metric.
As defined in ICARUS~\cite{giuliari2021icarus}, maxUp measures the maximum attack-induced traffic increase observed at any single source uplink.
Specifically, for each evaluated instance, HYDRA is constrained so that its maximum source uplink traffic does not exceed the ICARUS maxUp value.

\begin{table}[t]
\centering
\caption{Comparison between ICARUS and HYDRA under matched topology, traffic, routing, capacity, congestion-threshold, per-bot upload, and maxUp constraints.}
\label{tab:hydra_icarus_comparison}
\begin{tabular}{lrrr}
\hline
\textbf{Metric} & \textbf{ICARUS} & \textbf{HYDRA} & \textbf{Change} \\
\hline
Mean botnet size              & 1{,}096.56       & 725.18        & $-33.9\%$ \\
Median botnet size            & 1{,}012          & 652           & $-35.6\%$ \\
25th percentile               & 640              & 439           & $-31.4\%$ \\
75th percentile               & 1{,}542.5        & 1{,}039       & $-32.6\%$ \\
Mean attack flow (Mbps)       & 23{,}467.59      & 18{,}062.78   & $-23.0\%$ \\
Median attack flow (Mbps)     & 20{,}349         & 16{,}226      & $-20.3\%$ \\
\hline
\end{tabular}
\end{table}

As shown in Table~\ref{tab:hydra_icarus_comparison}, HYDRA reduces the required botnet size by 33.9\% on average relative to ICARUS.
This reduction is also consistent across the distribution, with the median botnet size reduced by 35.6\%, and the 25th and 75th percentile botnet sizes reduced by 31.4\% and 32.6\%, respectively.
HYDRA also reduces the aggregate attack flow by 23.0\% on average and 20.3\% at the median, indicating that the optimized allocation reduces both the number of active bots and the amount of attack traffic required to reach the modeled congestion condition.
These results show that HYDRA can achieve the same modeled attack objective as ICARUS while maintaining the same stealth threshold and requiring fewer resources.
While ICARUS completes its computation in seconds, HYDRA took around half a minute on average.
This runtime difference does not materially affect attack planning because HYDRA is used during the weaponization phase.

\subsection{Continuous Attack Evaluation}\label{sbsec:cont results}
In this evaluation, we assessed the continuous attack strategy described in \Cref{sec:cont attack}, executed over a 90-minute period with snapshots taken every 30 seconds.
This resulted in 180 snapshots, making joint optimization across all snapshots computationally challenging.
To examine the effect of the number of snapshots used for optimization, we selected evenly spaced subsets of snapshots from the 90-minute interval.
This process produced different botnets based on the selected snapshots.
We evaluated the effectiveness of each botnet by analyzing its ability to congest targeted paths.
This assessment was conducted for each 30-second snapshot.
Accuracy was measured as the percentage of successful attacks over the time window, representing the proportion of snapshots where the botnet achieved the intended congestion.

The computation time required to solve the optimization problem varied depending on the network conditions and targeted zones.
It ranged from a few seconds for simpler cases to a few hours for complex scenarios with larger numbers of snapshots, with an average computation time of 480 seconds.

\begin{figure}[hbt]
\centering
\includegraphics[width=\columnwidth]{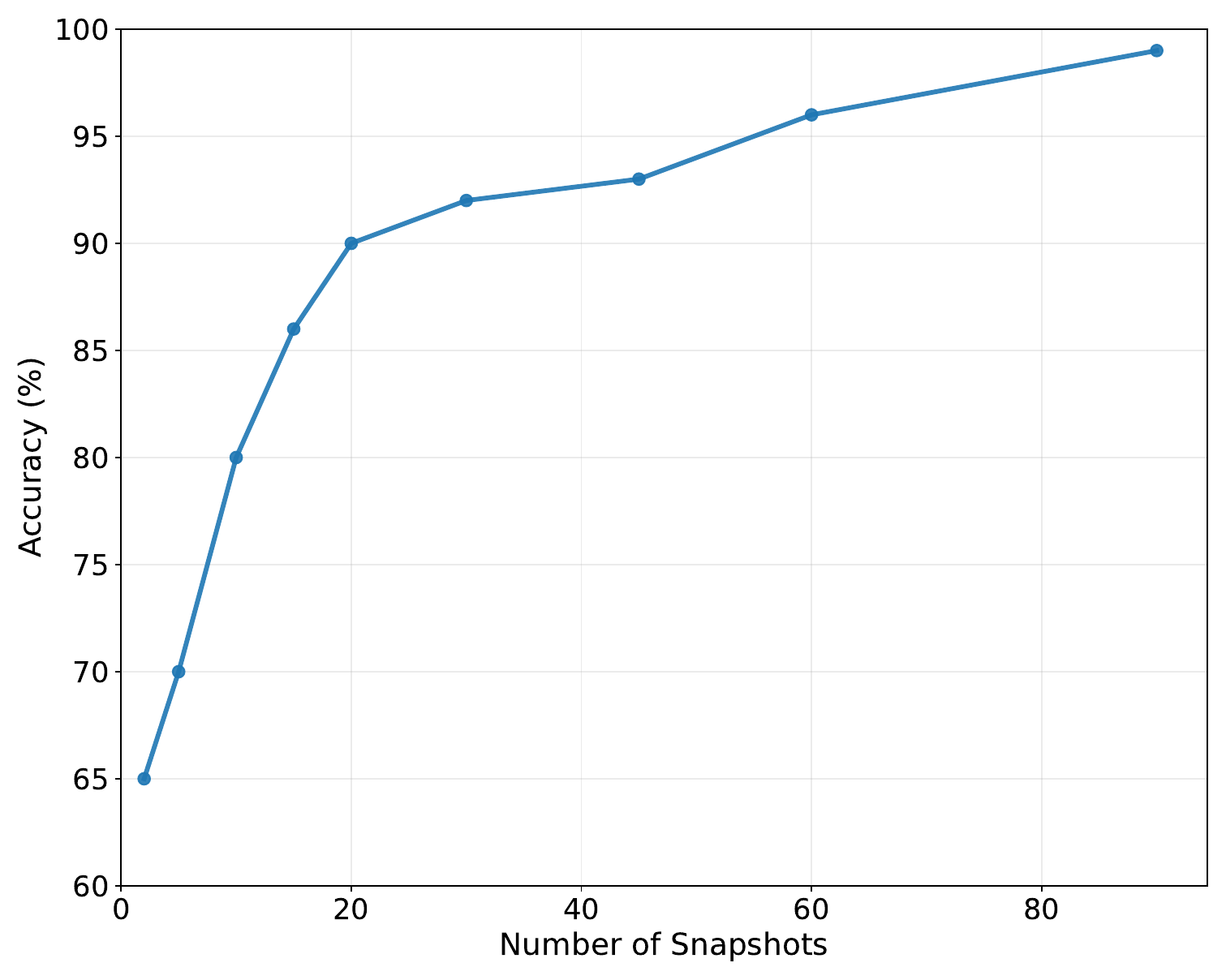}
\caption{Impact of the number of snapshots used for optimization on the attack success rate when performing continuous attacks over a 90-minute period.}
\label{fig:snapshot_accuracy_interval}
\end{figure}

\Cref{fig:snapshot_accuracy_interval} illustrates how the number of snapshots used in the optimization affects the success rate of continuous attacks over a 90-minute period. 
The botnet is obtained by solving the optimization problem with varying numbers of snapshots.
Using only 2--5 snapshots produces lower success rates.
As the snapshot count increases, performance surpasses 90\% with 30--60 snapshots, and approaches 99\% with 90 snapshots.
Although the use of fewer snapshots significantly reduces the computation time, continuous congestion is not sustained over the entire 90-minute period, indicating that higher snapshot counts enhance overall attack effectiveness.

To evaluate the attack strategy's robustness over an extended period, we used the botnet computed for one 90-minute interval with varying numbers of snapshots.
The scenario was analyzed over 90 hours with 30-second intervals, using the execution phase to assess attack accuracy over time.
The trend observed in \Cref{fig:snapshot_accuracy_interval} continued. 
Over the 90-hour evaluation period, the average success rate varied by only around 2\%, demonstrating the strategy's robustness and consistent high performance despite the extended duration and evolving network conditions.

\subsection{Flexible Botnet Configuration Attack Evaluation}\label{sbsec:flex results}

We evaluate the flexible botnet configuration approach, which is aimed at minimizing the size of the botnet while maintaining the ability to attack any selected zone pair, as described in \Cref{sec:felxible}.
Instead of solving each optimization problem independently, this approach identifies a global botnet, optimizing the allocation of resources across multiple attacks.
In this evaluation, we analyze the performance of the resulting attacks as a function of the number of targeted zone pairs.

\begin{figure}[hbt]
\centering
\includegraphics[width=\columnwidth]{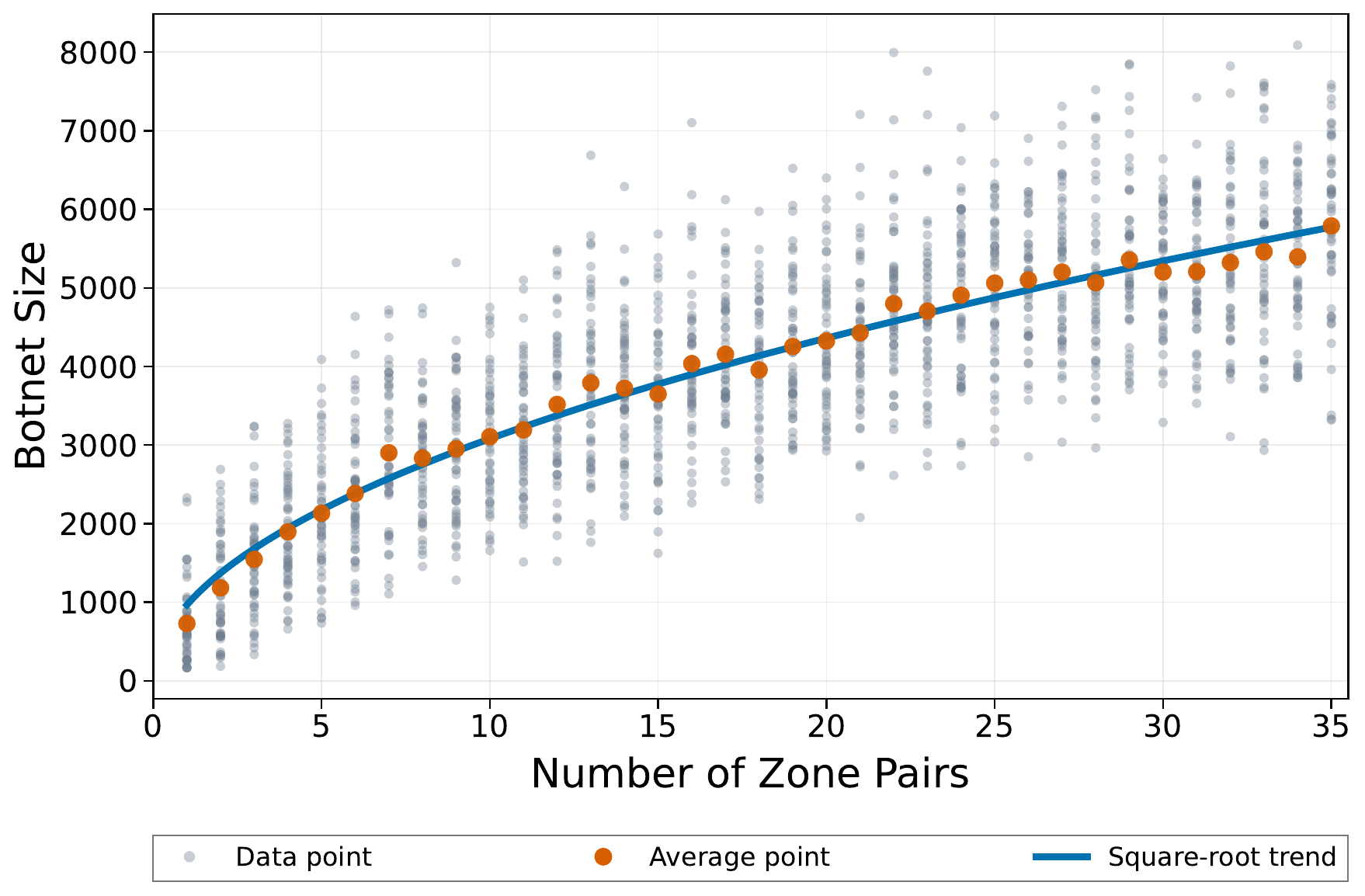}
\caption{Botnet size versus the number of targeted zone pairs. Gray points show individual runs, blue points show the means, and the red line shows the fitted square-root trend.}
\label{fig:graph_result_fig_4}
\end{figure}

\begin{figure}[hbt]
\centering
\includegraphics[width=\columnwidth]{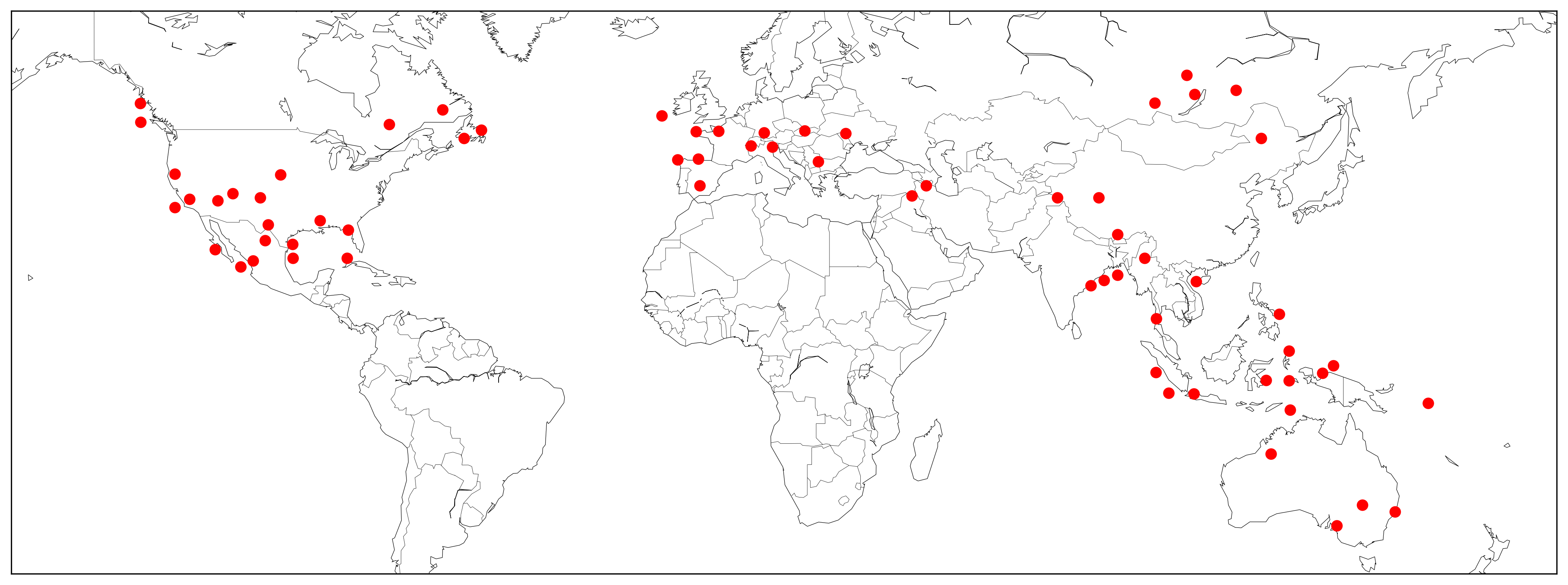}
\caption{Distribution of the botnet computed to attack 33 zone pairs, which achieved a 65\% success rate across 3,500 randomly sampled zone pairs.}
\label{fig:attack_indiv}
\end{figure}

For this analysis, we examined attacks targeting between 1 and 35 zone pairs.
The average botnet size required to attack each set of targeted zone pairs is presented in \Cref{fig:graph_result_fig_4}.
As the number of zone pairs increases, the size of the required botnet also increases.
However, the rate of growth decreases as the number of pairs increases.
This trend can be captured by a square root curve $y = 978\sqrt{x} - 12$, depicted in red, with $R^2 = 0.99$ in \Cref{fig:graph_result_fig_4}. The slowing growth rate in required botnet size demonstrates the efficiency of the flexible configuration.
This efficiency enables the targeting of more pairs with relatively few additional resources, which enhances the botnet's effectiveness in large-scale scenarios.
An example of the layout of the distribution of the computed botnet is illustrated in \Cref{fig:attack_indiv}, which depicts the computed layout for attacking 33 zone pairs.

\begin{figure}[hbt]
\centering
\includegraphics[width=\columnwidth]{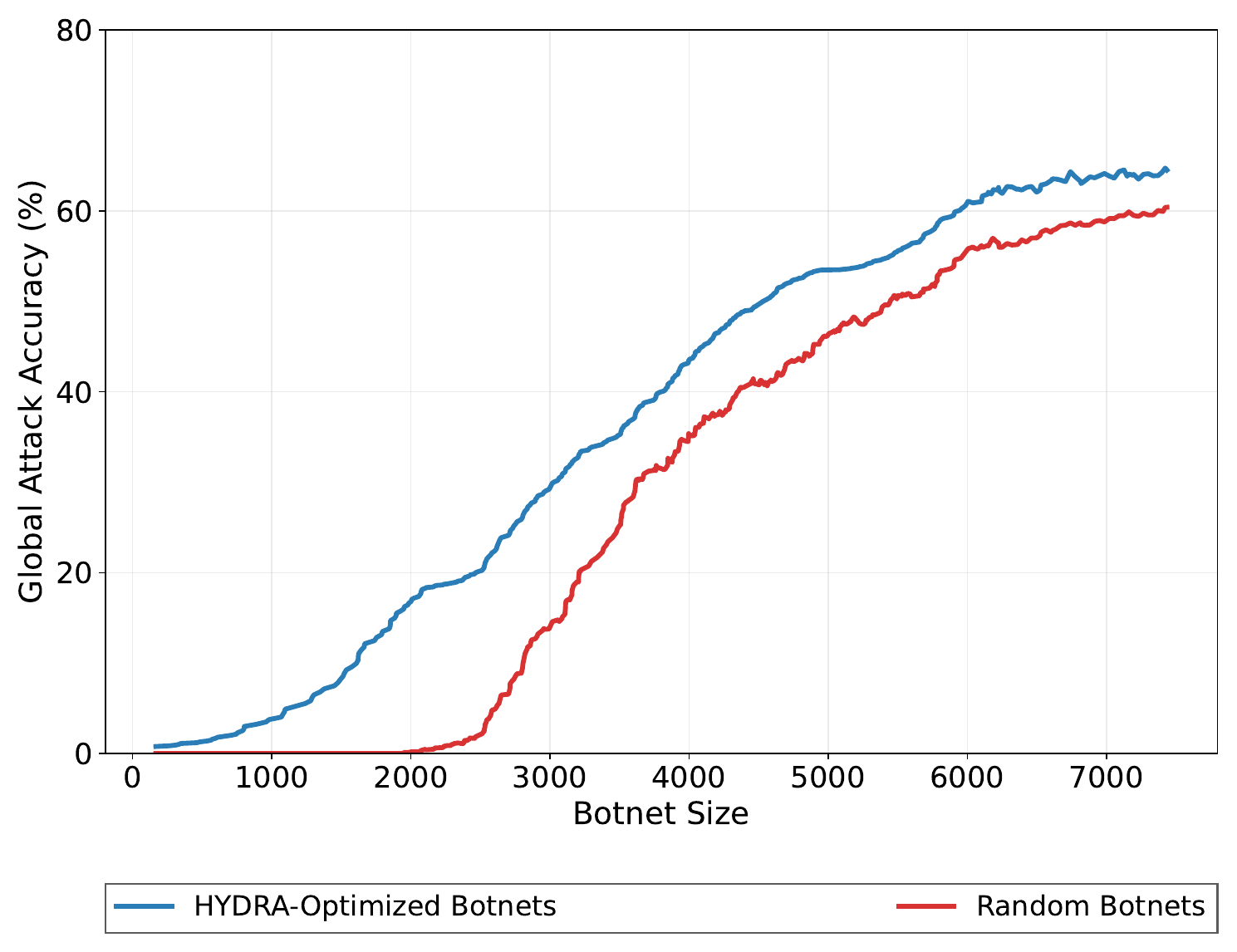}
\caption{Global attack accuracy as a function of botnet size for HYDRA-optimized botnets and randomly generated botnets of the same size.}
\label{fig:global_flex_attack}
\end{figure}

To assess the generalizability of the flexible configuration for global attacks, we randomly sampled 3,500 additional zone pairs based on GDP data and evaluated how effectively the different botnets could target them. 
The global attack accuracy, measured as the percentage of sampled pairs successfully attacked, is presented in \Cref{fig:global_flex_attack}.
The blue line represents the average global attack accuracy of HYDRA-optimized botnets weaponized in the flexible configuration phase described earlier in this subsection.
For comparison, we evaluated randomly generated botnets of the same size, which are represented by the red line.
For each HYDRA-optimized botnet size, we evaluated 10 different randomly generated botnets of the same size.
The results of these botnets were averaged to evaluate their effectiveness in attacking the sampled zone pairs.
HYDRA-optimized botnets outperformed the randomly generated ones, especially for smaller botnet sizes, where they achieved about 20\% higher global attack accuracy.
Even larger botnets maintained an advantage, attacking 5--10\% more zone pairs globally.
These results demonstrate the efficiency and scalability possible when strategically selecting botnet locations.

\subsection{Targeted Simultaneous Attacks}\label{sbsec:sim results}

In this section, we evaluate the simultaneous attack strategy described in \Cref{sec:Simul attack}, focusing on targeting multiple zone pairs concurrently.
We conducted the evaluation on groups of zone pairs, with each group containing 1--35 pairs; multiple evaluations were performed for each group size.
To assess this approach, we examined the botnet size as a function of the number of simultaneously attacked zone pairs.

\begin{figure}[hbt]
\centering
\includegraphics[width=\columnwidth]{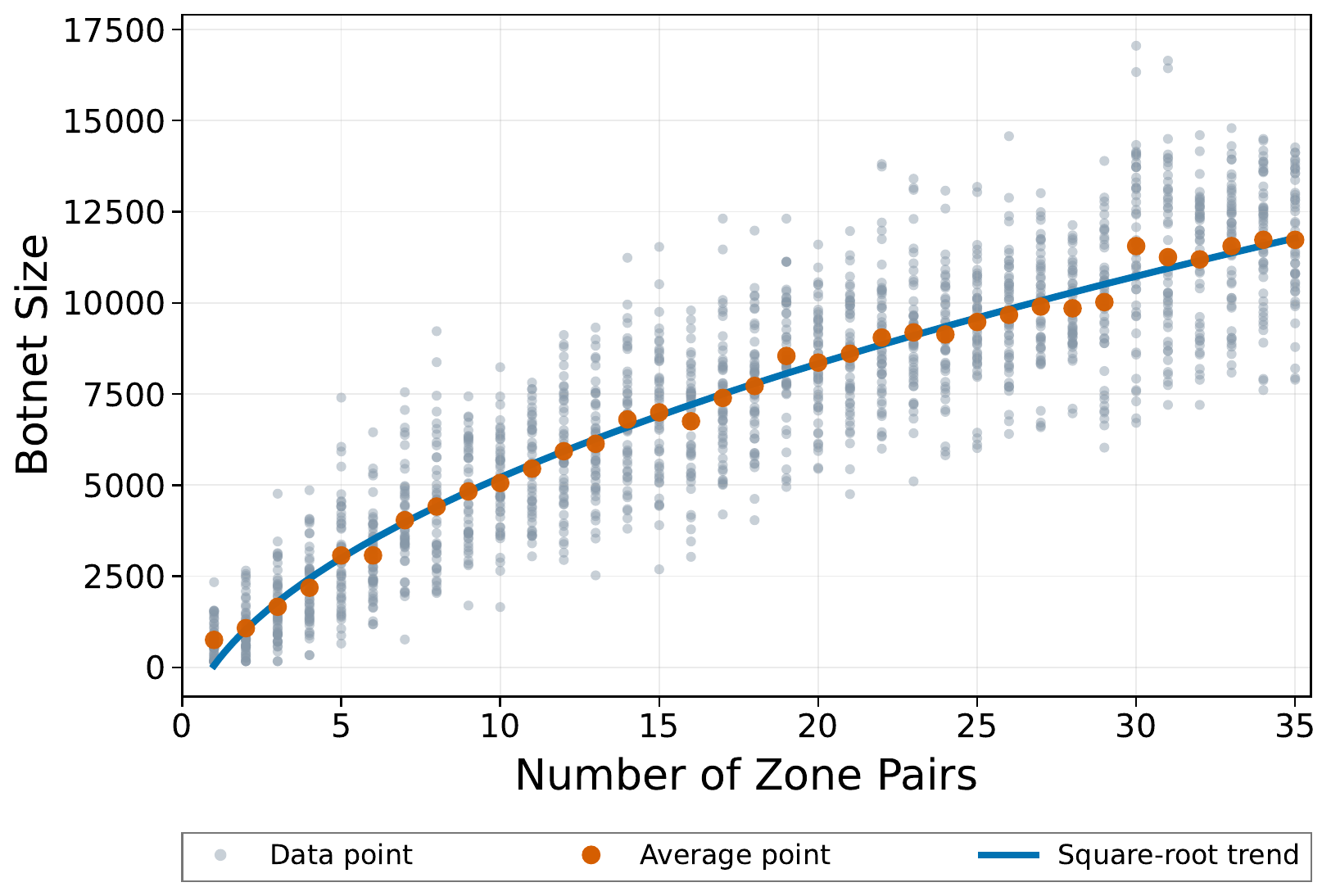}
\caption{Botnet size versus number of zone pairs simultaneously attacked.}
\label{fig:simul_fig}
\end{figure}

\Cref{fig:simul_fig} shows a strong square root trend in the relationship between botnet size and the number of simultaneously attacked zone pairs.
This trend can be captured by the curve $y = 2{,}385\sqrt{x} - 2{,}335$, with $R^2 = 0.992$, which is shown in red in \Cref{fig:simul_fig}.
This trend is largely influenced by zone pairs whose paths have ISLs in common or by attack paths that intersect multiple ISLs, which may become congested due to overlapping attacks.
Closer zone pairs are more likely to experience these effects, thereby reducing the required botnet size.

These results demonstrate efficient resource utilization, with square root growth in botnet size observed as the number of simultaneously attacked pairs increases.
For example, simultaneously attacking 15 pairs required an average botnet size of 6,994 bots.

\subsection{Sensitivity to Congestion and Bot Upload Parameters}
\label{sbsec:alpha_beta_sensitivity}

The main evaluation uses $\alpha=0.9$ and $\beta=25$~Mbps.
To assess whether the results depend on these specific values, we perform a sensitivity analysis over $\alpha \in \{0.85,0.90,0.95\}$ and $\beta \in \{10,25,50\}$~Mbps.
For each setting, we used the same topology, background traffic, routing model, and sampled zone pairs.
We report the percentage of feasible attacks and the mean required botnet size normalized to the corresponding $\beta=25$ setting for the same value of $\alpha$.

\begin{table}[t]
\centering
\caption{Attack sensitivity to $\alpha$ and $\beta$. Each cell reports feasibility percentage / mean botnet size relative to $\beta=25$ for the same $\alpha$.}
\label{tab:alpha-beta-all-links}
\begin{tabular}{c|ccc}
\hline
$\alpha$ & $\beta=10$ & $\beta=25$ & $\beta=50$ \\
\hline
0.85 & 89.73\% / 249.7\% & 89.68\% / 100.0\% & 89.71\% / 50.1\% \\
0.90 & 88.86\% / 249.7\% & 88.83\% / 100.0\% & 88.83\% / 50.1\% \\
0.95 & 87.74\% / 249.6\% & 87.73\% / 100.0\% & 87.76\% / 50.1\% \\
\hline
\end{tabular}
\end{table}

Table~\ref{tab:alpha-beta-all-links} shows two main trends.
First, increasing $\alpha$ slightly reduces feasibility, from about 89.7\% at $\alpha=0.85$ to about 87.7\% at $\alpha=0.95$.
This is expected because larger $\alpha$ values require links to be driven closer to saturation before they satisfy the congestion condition.
Second, changing $\beta$ has little effect on feasibility but changes the required botnet size almost proportionally.
Reducing the per-bot upload limit from 25~Mbps to 10~Mbps increases the mean required botnet size by about 2.5$\times$, while increasing it to 50~Mbps reduces the required botnet size by about half.
This indicates that $\beta$ primarily scales the number of bots needed to generate a feasible traffic allocation, whereas $\alpha$ affects whether the targeted congestion condition can be reached under the modeled topology and capacity constraints.
These results show that a network's resilience to LFAs depends on the upload capacity available to individual compromised terminals, since lower per-terminal upload capacities require larger botnets to achieve the same disruption.

\subsection{Cross-Constellation Sensitivity}

To assess whether HYDRA's feasibility trends are specific to the main Starlink-like shell, we evaluated single-snapshot attack instances across ten LEO constellation configurations.
For each configuration, we report the constellation parameters, the feasibility percentage, and the mean botnet size over feasible instances.

\begin{table*}[t]
\centering
\caption{Constellation sensitivity of the HYDRA minimum-user attack across representative LEO constellation scenarios.}
\label{tab:constellation_sensitivity}
\begin{tabular}{lrrrrrr}
\hline
\textbf{Scenario} & \textbf{Planes} & \textbf{Sats/plane} & \textbf{Incl.} & \textbf{Alt.} & \textbf{Feasibility} & \textbf{Mean bots} \\
 & & & \textbf{(deg.)} & \textbf{(km)} & & \\
\hline
Starlink baseline               & 72 & 22 & 53.0 & 550  & 85.8\% & 679 \\
Densified Starlink              & 72 & 30 & 53.0 & 550  & 92.3\% & 786 \\
OneWeb polar scenario           & 18 & 40 & 86.4 & 1200 & 95.6\% & 894 \\
Expanded OneWeb scenario        & 20 & 50 & 86.4 & 1200 & 99.6\% & 813 \\
Sun-synchronous scenario        & 34 & 32 & 97.4 & 600  & 99.6\% & 935 \\
Kuiper 630-km shell             & 34 & 34 & 51.9 & 630  & 97.2\% & 776 \\
Starlink 70$^\circ$ scenario    & 36 & 22 & 70.0 & 570  & 94.9\% & 884 \\
Kuiper 610-km shell             & 36 & 36 & 42.0 & 610  & 96.6\% & 1{,}047 \\
Low-inclination Walker scenario & 60 & 24 & 30.0 & 550  & 93.3\% & 1{,}045 \\
High-altitude Walker scenario   & 70 & 25 & 55.0 & 1200 & 89.3\% & 814 \\
\hline
\end{tabular}
\end{table*}

As shown in Table~\ref{tab:constellation_sensitivity}, HYDRA remains feasible across a broad range of LEO constellation configurations, with feasibility ranging from 85.8\% to 99.6\%.
The required botnet size also remains within the same order of magnitude across configurations, with mean botnet sizes ranging from 679 to 1{,}047 bots.
These results indicate that the observed attack feasibility is not limited to a single topology, although constellation structure affects both the percentage of feasible attack instances and the required botnet size.
\section{Mitigation Strategies}
\label{sec:mitigations}

In this section, we use HYDRA to evaluate how different mitigation mechanisms affect the resilience of LEO networks to LFAs.
Our goal is not to determine whether a single attack instance can be prevented, but to measure how each mechanism changes the adversary's overall ability to execute attacks at a global scale.
We define global attack success as the percentage of the 3,500 GDP-weighted randomly sampled zone pairs that a given attack configuration successfully attacks, using the same set of pairs from the flexible-botnet analysis.
This allows us to evaluate whether a mitigation reduces global attack success, increases the resources required for disruption, or limits the attacker's ability to reuse an optimized botnet across targets.

We evaluate five mitigation strategies that target different dependencies of the attack: routing diversification, link-triggered source throttling, ingress capacity policing, distance-based traffic constraints, and botnet attrition.
We then re-evaluate attack feasibility under the modified constraints and compare the resulting global attack success to the unmitigated baseline.
Potential QoS implications and operational trade-offs are discussed qualitatively at the end of this section.

\subsection{Route Diversification}

LFAs depend on routing behavior because the selected routes determine whether attack traffic reaches the links the adversary seeks to congest.
This makes routing a natural mitigation point: by diversifying route selection or avoiding heavily used links, the defender can reduce the attacker's ability to concentrate traffic on the same bottlenecks.
To evaluate this routing dependency, we implement three routing-based defenses from prior work: probabilistic routing~\cite{fratty2023random}, Bottleneck-Minimize Routing (KBM)~\cite{meng2025mitigating}, and segment-routing multipath inspired by landmark-based skeleton routing for broadband LEO constellations~\cite{hu2024lightweight}.
Probabilistic routing reduces path predictability, KBM avoids bottleneck-heavy routes, and segment-routing multipath increases route diversity.
Together, these mechanisms target the path consistency that LFAs exploit, making it harder for an adversary to direct limited botnet resources through the links selected for congestion.

Recent work on LEO routing has explored route selection as a defense mechanism against congestion and DDoS-style disruption.
For example, GRL-RR~\cite{bai2025grl} and STARCURE~\cite{lai2023achieving} adapt routing decisions to network conditions, while probabilistic routing~\cite{fratty2023random} introduces randomized path selection.
Following~\cite{fratty2023random}, for each source-destination pair the routing procedure randomly selects among four path-selection strategies with fixed probabilities: KSP with probability 0.703, KDG with probability 0.070, KDS with probability 0.143, and KLO with probability 0.084.
We then re-evaluate HYDRA using the routes produced by this randomized routing policy.

KBM follows the bottleneck-minimization routing approach of~\cite{meng2025mitigating}.
Each link is assigned a bottleneck score, and route computation uses this score as the Dijkstra edge weight rather than physical distance.
Candidate paths are accepted only if their physical length remains within a bounded stretch of the source-destination distance.
After a candidate path is selected, its ISL edges are disabled before searching for additional alternatives, encouraging path diversity and reducing repeated use of bottleneck links.
In our evaluation, we use a stretch bound of 1.53 and generate up to $k=5$ candidate paths.

We further evaluate segment-routing multipath inspired by LGSR~\cite{hu2024lightweight}.
LGSR partitions the constellation into regions, abstracts them into a skeleton graph, and uses segment routing to guide traffic across landmark-based skeleton paths.
Within each skeleton segment, probabilistic multipath forwarding spreads traffic over multiple local forwarding choices to improve load balancing and avoid hotspot congestion.
In our evaluation, we adapt this idea by generating $k=3$ segment-routed candidate paths per source-destination pair under a stretch bound of 1.70.

\begin{figure}[t]
\centering
\includegraphics[width=\columnwidth]{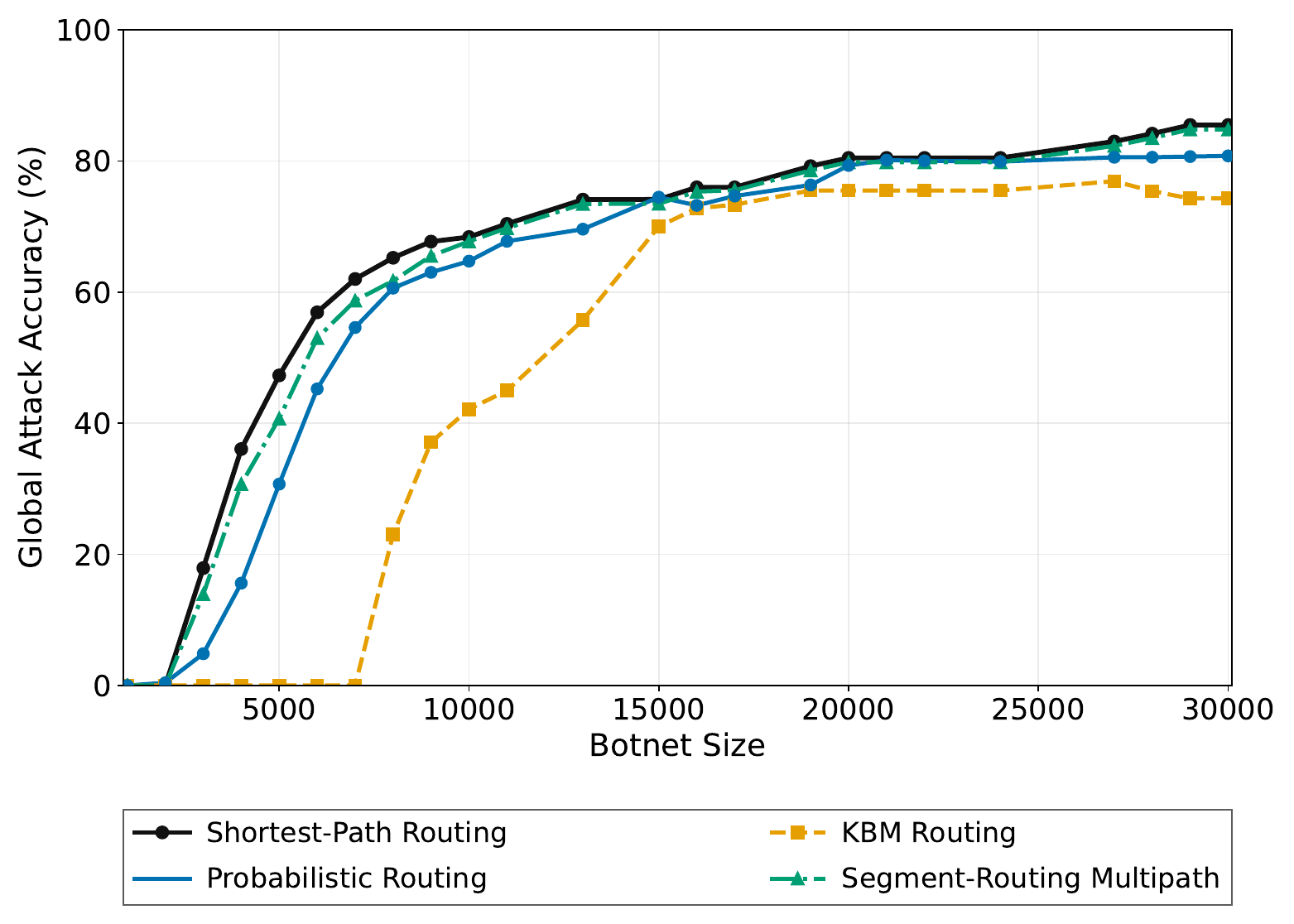}
\caption{Effect of route diversification on global attack success. Probabilistic routing reduces path predictability, KBM avoids bottleneck-heavy routes, and segment-routing multipath increases route diversity. KBM and probabilistic routing provide the strongest reductions relative to deterministic shortest-path routing, while segment-routing multipath provides a smaller reduction.}
\label{fig:probabilistic_routing_global_accuracy}
\end{figure}

As shown in \Cref{fig:probabilistic_routing_global_accuracy}, all three routing defenses reduce HYDRA's global attack success relative to shortest-path routing.
The reduction is most pronounced at small and medium botnet sizes.
KBM provides the strongest reduction, probabilistic routing produces a noticeable reduction, and segment-routing multipath provides a smaller but measurable reduction.
As the botnet size increases, the attacker partially recovers because a larger botnet provides more source-destination options for reaching targeted congestion links.
The multipath result shows that path diversity alone is not sufficient to fully neutralize HYDRA, since multiple candidate routes in the targeted area may still share exploitable bottlenecks or provide alternative bottlenecks that the attacker can target.
Overall, these results show that routing diversity and bottleneck-aware path selection can raise the botnet resources required for successful disruption, especially when the adversary operates with a limited bot population.

\subsection{Throttling Based on Congestion Source Attribution}

Next, we evaluate a mitigation strategy based on link-triggered source throttling.
This approach is inspired by prior DDoS defenses~\cite{ioannidis2002pushback,tan2007distributed} that treat overload as a congestion-control problem and apply rate limits to traffic aggregates that contribute to congestion.
When a link becomes congested, the defender applies rate limits to users or traffic aggregates that send traffic through the congested link.
This mitigation does not require identifying which users are malicious, because throttling is applied to all traffic sources associated with the congested link.

Let $\mathcal{U}_{t,l}$ denote the set of users or traffic aggregates whose traffic traverses link $l$ at snapshot $t$.
For each user $u \in \mathcal{U}_{t,l}$, let $\kappa_u^{(r)}$ be the effective sending limit after throttling round $r$.
The total traffic sent by user $u$ is constrained by

\begin{equation}
    R_{t,u} \leq \kappa_u^{(r)},
\end{equation}
where $R_{t,u}$ denotes the aggregate traffic generated by user $u$ at snapshot $t$.
After each throttling round, users associated with congested links receive a lower sending limit.
We evaluate two simple policies.
The first applies a hard cap

\begin{equation}
    \kappa_u^{(r+1)}
    =
    \min \left(
    \kappa_u^{(r)},
    \kappa_{\mathrm{cap}}
    \right).
\end{equation}

The second reduces the limit gradually by a fixed step while preserving a minimum floor

\begin{equation}
    \kappa_u^{(r+1)}
    =
    \max \left(
    \kappa_{\min},
    \kappa_u^{(r)} - \Delta
    \right).
\end{equation}
We then re-evaluate HYDRA under these updated source-side bandwidth limits.

\begin{figure}[t]
\centering
\includegraphics[width=\columnwidth]{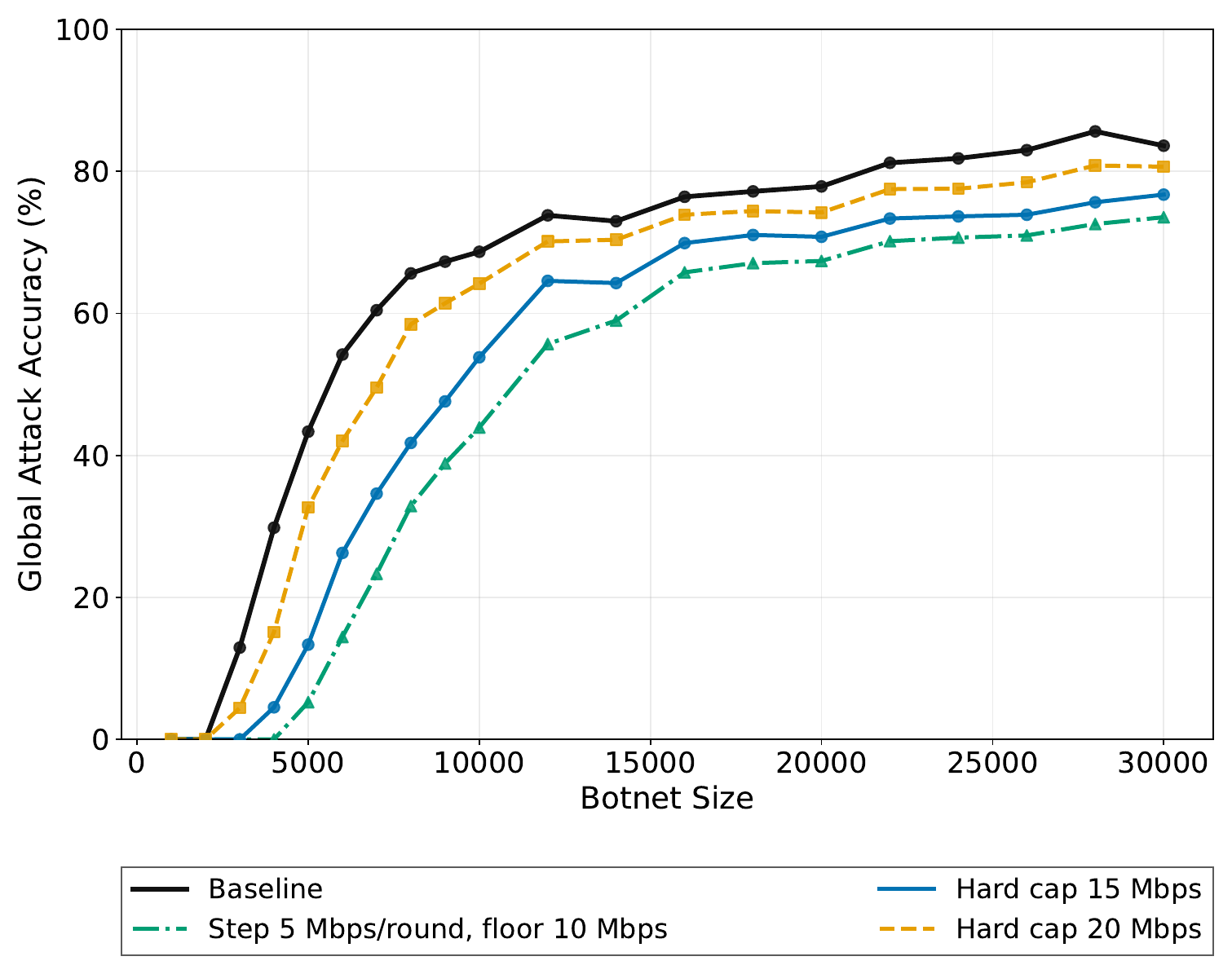}
\caption{Effect of link-triggered source throttling on global attack success. Throttling reduces the bandwidth available to traffic sources associated with congested links, either through a hard cap or a fixed-step reduction.}
\label{fig:link_throttling_global_accuracy}
\end{figure}

Figure~\ref{fig:link_throttling_global_accuracy} shows that link-triggered source throttling reduces HYDRA's attack success relative to the baseline.
The effect is most visible at small and medium botnet sizes, where reducing the bandwidth of congestion-contributing sources leaves the attacker with fewer effective allocation options.
This effect is strongest under repeated throttling, because sources that continue to contribute to congestion have their effective sending limits reduced over multiple rounds.
At larger botnet sizes, the attacker has more sources available, so the relative benefit of throttling becomes smaller.

\subsection{Ingress Capacity Policing}

Another strategy we evaluate is ingress capacity policing, a mitigation that limits the traffic admitted into the constellation at network entry points.
This idea follows traffic conditioning and ingress policing, where traffic is regulated close to where it enters the network~\cite{blake1998architecture,heinanen1999two}.
In LEO satellite networks, uplink GSLs provide natural enforcement points because user traffic enters the constellation only through these links.
The geographic origin of uplink traffic can also be estimated from physical-layer measurements, as prior work shows that satellite constellations can localize ground devices using received signal strength and Doppler measurements~\cite{hashim2022satellite}.
By enforcing admission budgets at the uplink GSLs or at geographic source areas, the defender can constrain excess traffic before it propagates through the network and contributes to downstream congestion.
This is especially relevant when adversarial traffic is concentrated in specific geographic regions, since ingress policing limits the excess traffic admitted from those regions.
These admission budgets should reflect the expected nominal demand of each region, so that regular service is preserved while unusually large traffic injections are limited.
We evaluate this mitigation using fixed caps on uplink GSL admission capacity and demand-based caps on the upload traffic admitted from each source area.

To model fixed ingress caps at the GSL level, let $\rho \in (0,1]$ denote the ingress-cap ratio.
Under ingress capacity policing, the total traffic admitted through each GSL is limited to a fraction $\rho$ of that link's capacity
\begin{multline}
\sum_{\substack{g_s,g_d:\\ l \in \mathcal{F}_t(g_s,g_d)}}\hspace*{-1.2em}
\left(\mathbf{TM}_{t,g_s,g_d}+\mathbf{ATM}_{t,g_s,g_d}\right)
\leq \rho c_l^{\max}, \\
\quad \forall l \in L^{GSL}_t .
\end{multline}

For the demand-based source-area policy, each GPO serves as a source area.
Let $\bar{\lambda}_g$ denote the expected nominal upload demand from source area $g \in N^{GPO}$, estimated from the background traffic profile, and let $\eta$ denote the allowed excess-demand ratio.
The resulting constraint limits the total traffic admitted from each source area relative to its expected nominal demand as
\begin{equation}
\sum_{g_d}
\left(\mathbf{TM}_{t,g,g_d}+\mathbf{ATM}_{t,g,g_d}\right)
\leq (1+\eta)\bar{\lambda}_g,
\forall g \in N^{GPO}.
\end{equation}

\begin{figure}[t]
\centering
\includegraphics[width=\columnwidth]{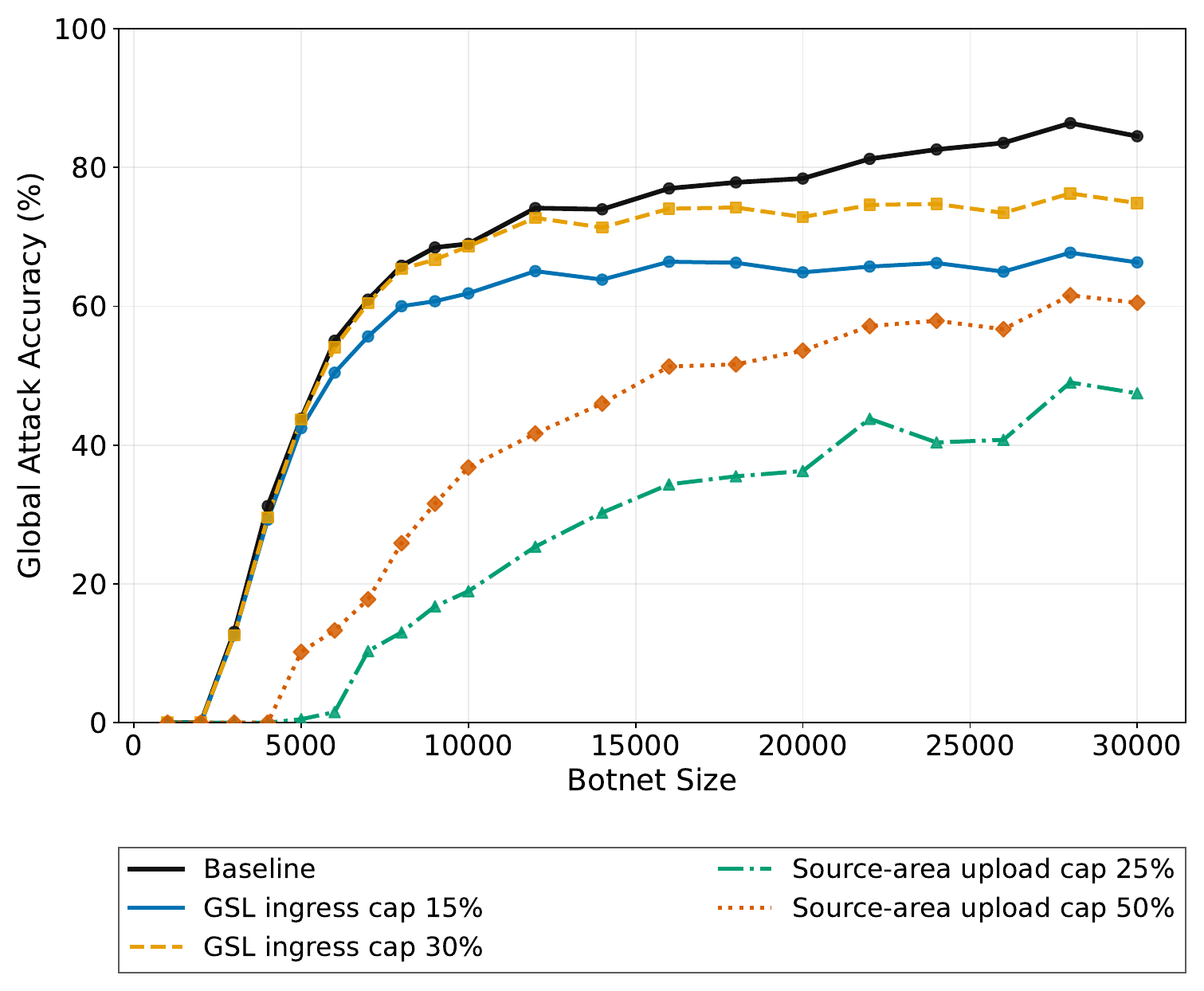}
\caption{Effect of ingress capacity policing on global attack success. GSL ingress caps limit uplink admission capacity directly, while source-area upload caps limit admitted traffic relative to expected nominal demand from each source area.}
\label{fig:ingress_capacity_policing}
\end{figure}

Figure~\ref{fig:ingress_capacity_policing} shows that ingress policing reduces HYDRA's global attack success, but the effect differs across the two policies.
Source-area upload caps produce the strongest reduction, especially as the botnet size increases.
This occurs because larger botnets may place more bots in the same source areas, where their combined upload is constrained by the same demand-based cap, limiting the usable contribution of additional bots in those areas.
GSL ingress caps also reduce attack success by limiting the total traffic admitted through each uplink link, but their effect is less pronounced because each cap applies to aggregate GSL traffic rather than to the upload budget of a specific source area.
Overall, these results show that ingress policing can hinder LFA-style attacks in satellite constellations and increase the resources required for successful disruption.

\subsection{Distance-Based Traffic Constraints}

Distance-based traffic constraints limit the amount of traffic that can be sent between geographically distant source-destination pairs.
The mitigation applies stricter sending limits as the distance between the source and destination increases, while leaving shorter flows less affected.
For LFAs, this reduces the attacker's ability to use distant destinations to steer traffic from distributed bot locations through targeted congestion links.

Let $d(g_s,g_d)$ denote the geographic distance between GPOs $g_s$ and $g_d$.
We model distance-based mitigation as a per-user sending limit that depends on the distance to the destination GPO.
For each user located at $g_s$ and sending traffic toward $g_d$, the admissible sending rate is capped by $\sigma_{t,g_s,g_d}\beta$, where $\sigma_{t,g_s,g_d}\in[0,1]$ and $\beta$ is the nominal per-user sending limit.
When $\sigma_{t,g_s,g_d}=1$, the user's sending limit is unchanged and $\sigma_{t,g_s,g_d}=0$ blocks that source-destination direction.
We evaluate several forms of $\sigma_{t,g_s,g_d}$. The first is a hard distance limit

\begin{equation}
    \sigma_{t,g_s,g_d}
    =
    \begin{cases}
    1, & d(g_s,g_d) \leq d_{\max},\\
    0, & d(g_s,g_d) > d_{\max}.
    \end{cases}
\end{equation}
This policy blocks source-destination flows whose geographic distance exceeds $d_{\max}$. 
We also evaluate a gradual distance-decay policy, where the allowed sending rate decreases as the source-destination distance increases beyond a reference distance $d_0$

\begin{align}
    &\sigma^{\mathrm{exp}}_{g_s,g_d}
    =\notag \\
    &\begin{cases}
    1, & d(g_s,g_d) \leq d_0,\\
    \max \left(
    \sigma_{\min},
    e^{
    -\ln(2)
    \frac{d(g_s,g_d)-d_0}{d_{1/2}}}
    \right), & d(g_s,g_d)>d_0.
    \end{cases}
\end{align}
For source-destination pairs with $d(g_s,g_d) \leq d_0$, no rate reduction is applied.
For pairs with $d(g_s,g_d)>d_0$, the allowed sending rate decreases with distance until it reaches the minimum value set by $\sigma_{\min}$.
The parameter $d_{1/2}$ controls how quickly this decrease occurs.

We also consider a hop-aware traffic constraint, where route length is represented by the number of satellites traversed by a source-destination flow.
In this mitigation, source-destination pairs that traverse more satellites receive a lower sending limit.
Let $h(g_s,g_d)$ denote the representative number of satellites traversed by traffic from $g_s$ to $g_d$, assuming the shortest available route.
For each GPO pair, we define the hop-aware sending-rate factor as
\begin{equation}
    \sigma^{\mathrm{hop}}_{g_s,g_d}
    =
    \begin{cases}
    1, & h(g_s,g_d) \leq h_{\mathrm{free}},\\
    \max \left(
    \sigma_{\min},
    \left(
    \frac{h_{\mathrm{free}}}{h(g_s,g_d)}
    \right)^q
    \right), & h(g_s,g_d)>h_{\mathrm{free}},
    \end{cases}
\end{equation}
where $h_{\mathrm{free}}$ is the number of hops allowed before throughput reduction is applied, $q$ controls the aggressiveness of the hop-based throughput reduction, and $\sigma_{\min}$ is the minimum permitted sending-rate multiplier.
Thus, routes that traverse more satellites receive a lower per-user sending limit.

\begin{figure}[t]
\centering
\includegraphics[width=\columnwidth]{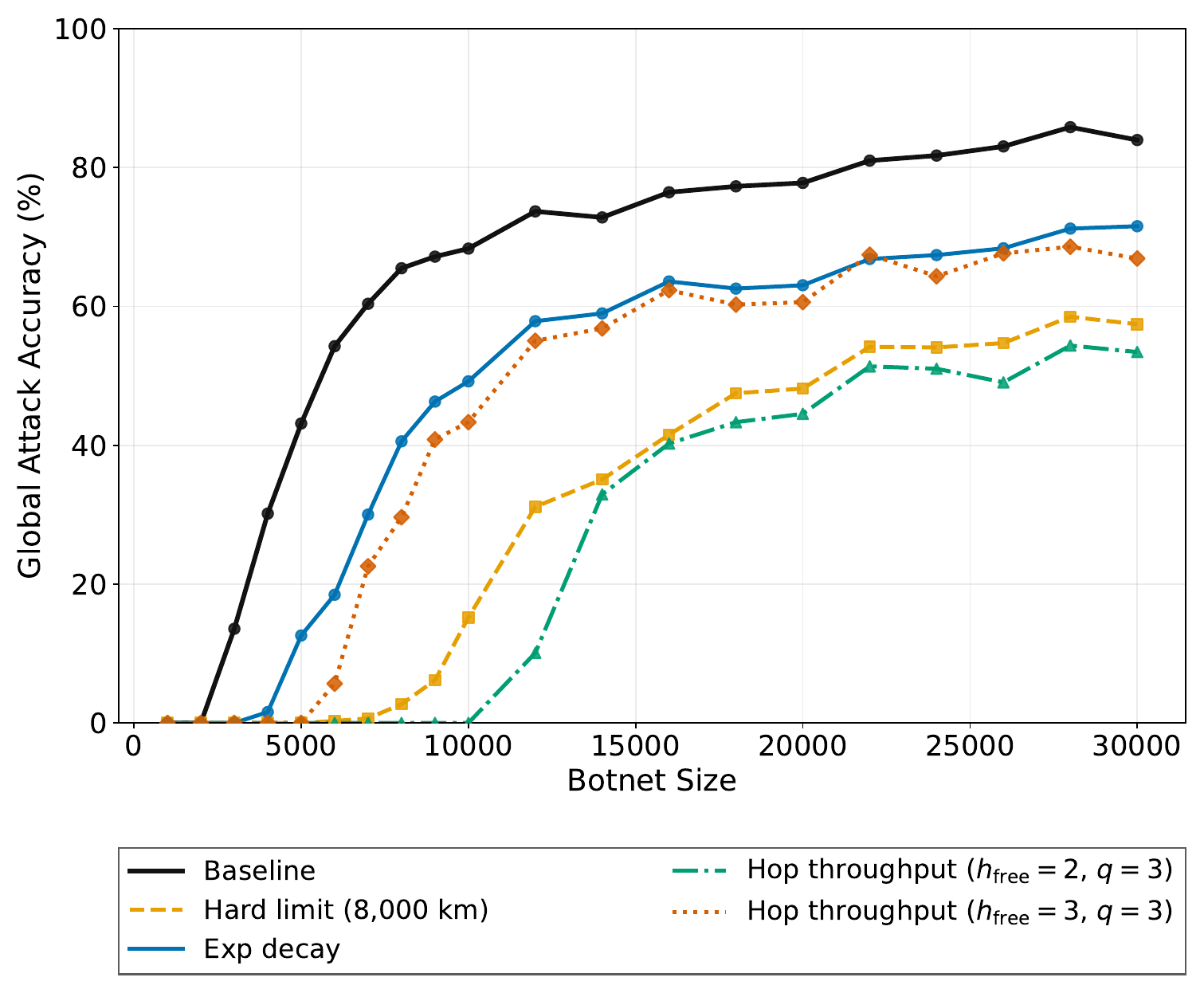}
\caption{Effect of distance-based and hop-aware traffic constraints on global attack success. Hard distance limits block long-distance source-destination flows, exponential decay gradually reduces admissible traffic with geographic distance, and hop-throughput policies reduce admissible traffic for routes that traverse more satellites.}
\label{fig:max_distance_global_accuracy}
\end{figure}

Figure~\ref{fig:max_distance_global_accuracy} shows that distance-based and hop-aware traffic constraints reduce HYDRA's global attack success relative to the baseline.
The strongest reduction is obtained by the hop-throughput policy with $h_{\mathrm{free}}=2$ and $q=3$, which sharply limits the contribution of longer satellite-hop routes and keeps attack success substantially below the baseline across botnet sizes.
The hard 8,000 km limit also provides a strong reduction by removing long-distance source-destination flows entirely.
Softer policies, such as exponential distance decay and the hop-throughput policy with $h_{\mathrm{free}}=3$, reduce attack success more gradually, but allow the attacker to recover more as botnet size increases.
These results show that both geographically distant flows and longer satellite-hop routes provide useful attack options for reaching bottleneck links.
Constraining these flows reduces the feasible attack space and increases the resources required for successful disruption.

\subsection{Botnet Attrition}

We next evaluate botnet attrition as a reactive mitigation that removes attacker-controlled bots after attack activity is observed.
This models remediation actions such as blocking, cleanup, quarantine, or loss of attacker control over participating bots.
Since HYDRA repeatedly relies on effective bot locations, removing used bots can force the attacker to re-optimize with a smaller and less favorable bot pool.

Let $\mathcal{B}^{(r)}$ denote the set of bots available to the attacker after attrition round $r$, and let $\mathcal{U}^{(r)} \subseteq \mathcal{B}^{(r)}$ denote the bots used by the optimized attack allocation in that round.
Let $\mathcal{A}_{\eta}$ denote an attrition process with attrition
intensity $\eta \in [0,1]$. After round $r$, the defender removes a
subset of the bots used by the optimized attack allocation,

\begin{equation}
    \mathcal{R}^{(r)}
    =
    \mathcal{A}_{\eta}\left(\mathcal{U}^{(r)}\right),
    \qquad
    \mathcal{R}^{(r)} \subseteq \mathcal{U}^{(r)} .
\end{equation}

The attacker pool for the next round is then updated as

\begin{equation}
    \mathcal{B}^{(r+1)}
    =
    \mathcal{B}^{(r)} \setminus \mathcal{R}^{(r)} .
\end{equation}
Equivalently, each removed bot $b \in \mathcal{R}^{(r)}$ has zero effective sending capacity in subsequent evaluations

\begin{equation}
    \beta_b^{(r+1)} = 0 .
\end{equation}

\begin{figure}[t]
\centering
\includegraphics[width=\columnwidth]{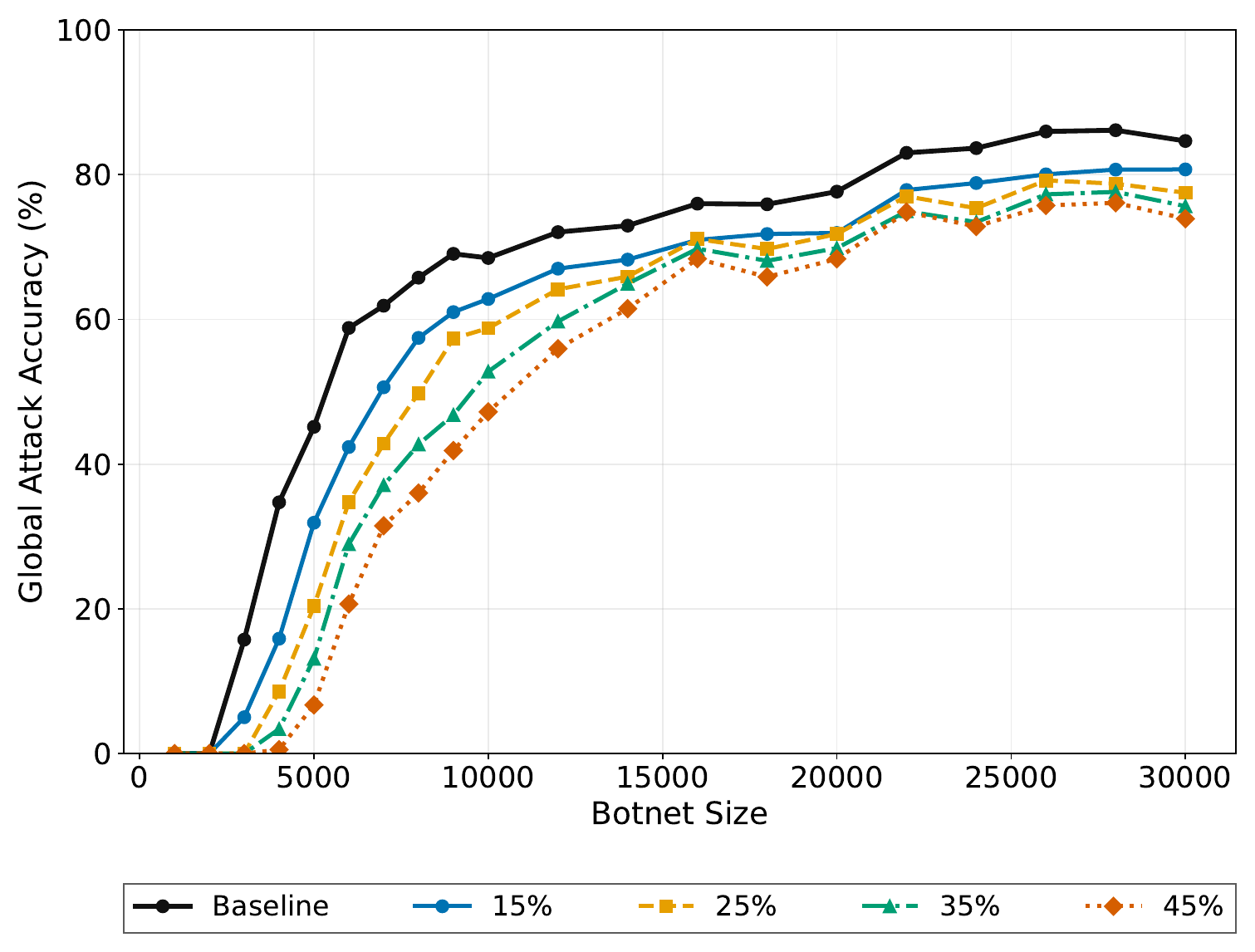}
\caption{Impact of botnet attrition on global attack success. Higher attrition rates reduce the attacker's ability to re-optimize after used bots are removed from the available bot pool.}
\label{fig:botnet_attrition_global_accuracy}
\end{figure}

In our experiments, we instantiate $\mathcal{A}_{\eta}$ by uniformly sampling an $\eta$ fraction of the bots used by the optimized attack allocation in each attrition round.
We evaluate attrition rates of 15\%, 25\%, 35\%, and 45\%.
Figure~\ref{fig:botnet_attrition_global_accuracy} shows that botnet attrition reduces HYDRA's attack success relative to the baseline, with the strongest impact at small and medium botnet sizes.
HYDRA minimizes the active bot resources required for each attack, so attrition is applied to the compact set of bots exposed during the optimized attack rather than to the entire candidate botnet.
When the available bot pool is small, removing part of this active set leaves fewer alternatives for re-optimization, leading to a larger reduction in attack feasibility.
For larger botnets, many candidate bots remain unused during each optimized attack, allowing the attacker to replace removed bots from reserve sources in subsequent iterations.

These results suggest that attrition is most effective when the attacker has limited reserve capacity.
Combining attrition with the mitigation strategies evaluated earlier is therefore a more promising direction for reducing the feasibility of HYDRA and similar LFA attacks.

\subsection{Operational Deployment Trade-offs}

Implementing mitigation mechanisms in real LEO networks introduces a trade-off between reducing attack feasibility and preserving quality of service (QoS). 
Stricter mitigation policies can make LFAs harder to execute, but they may also reduce QoS by increasing latency, limiting legitimate traffic bursts, or degrading service for users whose traffic matches the mitigation policy.

Route diversification can reduce path predictability, but longer or less direct routes may increase latency.
Link-triggered source throttling can relieve congested links, but it may also rate limit benign users whose traffic traverses those links.
Ingress capacity policing can constrain excessive uplink traffic at network entry points, but strict admission budgets may reduce service quality during legitimate demand spikes.
Distance-based traffic constraints can limit the attacker's use of globally distributed bots, but distance-aware rate shaping may also degrade QoS for legitimate long-distance traffic, a core use case of satellite constellations.
Botnet attrition can weaken the attacker's available bot pool, but aggressive remediation may disrupt service for legitimate users if detection misidentifies benign activity as malicious.

The goal of this mitigation analysis is to show the defensive potential of these directions and quantify how they strengthen the network's resilience to LFAs.
In practice, each mitigation would need to be tailored to the operator's constellation design, routing architecture, customer base, service-level requirements, and acceptable QoS impact.
A production deployment would therefore require operator-specific tuning of enforcement thresholds, traffic policies, and service guarantees.
\section{Conclusion}\label{sec:conclusions}

This paper introduced HYDRA, an optimization-based framework for modeling strategic LFAs in dynamic LEO satellite networks.
HYDRA captures the attack planning problem under time-varying topology, limited bot availability, per-bot upload constraints, and geographic placement constraints.
Rather than assuming widespread control over compromised devices, as in prior work, HYDRA identifies the smallest active bot set and traffic allocation needed to congest GSL and ISL links between targeted geographic zones across time-varying topologies.

Our evaluation shows that HYDRA provides a resource-threshold view of LEO network resilience to LFAs by quantifying the minimum botnet resources required for targeted disruption.
In single-snapshot zone-attack instances, HYDRA reduces the active botnet required to reach the same modeled congestion objective by 34\% relative to an ICARUS stealth-distribution baseline under matched detectability constraints, while also reducing aggregate attack flow by 23\%.
Under the modeled routing and traffic assumptions, HYDRA maintains over 97\% success in continuous attacks and scales efficiently to multiple targets, with square-root botnet growth as the number of targeted zone pairs increases.
We also show the defensive potential of routing diversification, source throttling, ingress policing, distance-based traffic constraints, and botnet attrition in strengthening the network's resilience to LFAs.

These findings highlight structural vulnerabilities in LEO constellations and reinforce the need for defenses that account for routing behavior, traffic admission, source bandwidth limits, botnet reuse, and the constellation's evolving topology.
As LEO satellite networks continue to expand and support increasingly important communication services, understanding the resource requirements of LFAs and the trade-offs involved in mitigating them is essential for improving network resilience against such attacks.

\section*{Ethical Considerations}

This work analyzes LFAs in LEO satellite networks to quantify adversarial resource thresholds and evaluate defensive strategies.
The study is conducted entirely in simulation and does not involve experiments on operational satellite systems, production networks, or real users. 
HYDRA is intended as a risk-assessment and resilience-evaluation framework for researchers and network operators, enabling them to assess how routing, capacity controls, throttling, traffic constraints, and botnet availability affect attack feasibility.
Because the techniques studied in this paper are dual-use, we frame the analysis around defensive planning, capacity assessment, and mitigation evaluation rather than operational guidance for disrupting deployed satellite services.
\section*{Acknowledgment}
The authors used OpenAI's ChatGPT and Anthropic's Claude to assist with grammar review, language review, and editorial refinement throughout the manuscript.
The tools were used to improve clarity and presentation, not to generate experimental results, figures, or technical claims.
The authors reviewed and edited all assisted text.

\bibliographystyle{IEEEtran}
\bibliography{references}

\end{document}